\documentclass{aa}  
\usepackage{graphicx}
\usepackage{bm}

\usepackage{txfonts}
\begin{document}

   \title{Characterizing the line emission from molecular clouds }
   \subtitle {II. A comparative study of California, Perseus, and Orion A}  
   
   \titlerunning{A comparative study of California, Perseus, and Orion A}
   
    \author{M. Tafalla\inst{1} \and A. Usero\inst{1} \and A. Hacar\inst{2}}
    
    \institute{Observatorio Astron\'omico Nacional (IGN), Alfonso XII 3,
       E-28014 Madrid, Spain \\
    \email{m.tafalla@oan.es, a.usero@oan.es}
         \and
     Department of Astrophysics, University of Vienna, T\"urkenschanzstrasse 17, 
     1180 Vienna, Austria \\
     \email{alvaro.hacar@univie.ac.at} }

   \date{Received ; accepted }

 
\abstract
{}
{We aim to characterize and compare the molecular-line emission of three clouds whose
star-formation rates span
one order of magnitude: California, Perseus, and Orion A.}
{We used stratified random sampling to select
positions representing the different
column density regimes of each cloud and
observed them with the IRAM 30 m telescope.
We covered the 3 mm wavelength band and focused
our analysis on CO, HCN, CS, HCO$^+$, HNC, and N$_2$H$^+$.}
{We find that the line intensities depend most strongly
on the H$_2$ column density, with which they are tightly correlated.
A secondary effect, especially visible in Orion A, is a dependence
of the line intensities on the gas temperature.
We explored a method that corrects for temperature variations and show that,
when it is applied, the emission from the three clouds behaves very similarly.
CO intensities vary
weakly with column density, while the intensity of
traditional dense-gas tracers
such as HCN, CS, and HCO$^+$ varies almost linearly with column density.
N$_2$H$^+$  differs from all other species in that it traces only cold
dense gas.
The intensity of the rare HCN and CS isotopologs reveals additional
temperature-dependent abundance variations.
Overall, the clouds have similar
chemical compositions 
that, as the depth increases, are sequentially 
dominated by photodissociation, gas-phase reactions, molecular freeze-out, and stellar feedback in the densest parts of Orion A.
Our observations also allowed us to calculate line luminosities for each cloud,
and a comparison with literature values shows good agreement.
We used our HCN(1--0) data to explore the behavior of the HCN conversion factor,
finding that it is dominated by the emission from the outermost cloud
layers. It also depends strongly on the gas kinetic temperature.
Finally, we show that the
HCN/CO ratio provides a
gas volume density estimate, and that its correlation with the
column density resembles that found in extragalactic
observations.}
{}


\maketitle
%

\section{Introduction}
\label{sec_intro}

Characterizing the large-scale emission of molecular clouds
is necessary to determine their internal structure and star-formation properties and
help connect galactic and extragalactic observations.
It is, however, a challenging task due to the large size of the
clouds and the limited bandwidth and pixel number
of heterodyne receivers.
The first efforts to characterize the molecular emission from full clouds
focused on mapping the bright lines of CO and its isotopologs
(e.g., \citealt{ung87,dam01,rid06,gol08}).
CO, however, is easily thermalized, so its emission
is insensitive to the different density regimes of a cloud.
In the past decade,
a new generation of wide-band heterodyne receivers has
made it possible to observe multiple lines simultaneously \citep{car12},
and this has led to a new generation of multi-tracer studies of molecular clouds
\citep{kau17,pet17,shi17,wat17,bar20}.
These studies have characterized cloud emission by making fully sampled
maps, a technique that provides a very detailed picture of the gas distribution
but often requires hundreds of hours of telescope time.
For this reason, multiline studies have been
restricted to single clouds (or
parts of them), making it very difficult to compare the
emission of different targets.

While maps provide the detailed description required
to characterize the distribution
of cloud material into filaments and cores, or determine its velocity
field, clouds are turbulent objects whose structure is expected to be
mostly transient \citep{lar81,hey04}.
It is therefore likely that many properties of a cloud
do not depend on the small-scale details
of its emission and can be captured without the need of making maps.
Following this idea, \citet[hereafter Paper I]{taf21} presented
an alternative method for characterizing the emission of a molecular cloud
by observing a limited number of positions selected using
stratified random sampling.
This technique is commonly used in polling \citep{coc77}
and selects the positions to be observed by
first dividing the cloud into a number of column density bins and
then choosing a set of random target positions from each bin.
To apply this technique to the Perseus molecular cloud,
Paper I used the column density map presented by \cite{zar16} and
divided the cloud into ten logarithmically spaced
H$_2$ column density bins. From each bin, a set of ten cloud
positions were chosen at random, creating
a sample of 100 target positions that were observed with the
Institut de Radioastronomie Millimétrique (IRAM) 30 m telescope.

An advantage of the stratified sampling technique
is that it requires significantly less telescope time than mapping,
allows deep integrations to be obtained at low column densities,
and, as shown in Paper I, can accurately estimate basic emission
properties of a cloud, such as
the mean intensity and its dispersion inside each column density bin.
These parameters can later be compared with the results from
numerical simulations to test models of cloud formation \citep{pri23}.

In this paper we
present the results obtained from sampling the emission of the California and
Orion A clouds using the stratified random sampling technique, and we
compare the results with those of the Perseus cloud already presented
in Paper I.
California and Orion A are highly complementary to Perseus because they
are also nearby but are forming stars at very different rates. Distances to
California, Perseus, and Orion A have been estimated as $470\pm24$~pc,
$294\pm15$~pc, and $432\pm22$~pc,
respectively by \cite{zuc19} using \textit{Gaia} Data Release 2 data, although these quantities
should be considered as mean values given the complex 3D morphology of each
cloud \citep{gro18,rez22}. Concerning their star-formation
activity, \cite{lad10} estimated star-formation rates that span one
order of magnitude: 70~M$_\sun$ Myr$^{-1}$ for California,
150~M$_\sun$ Myr$^{-1}$ for Perseus, and 715~M$_\sun$ Myr$^{-1}$ for Orion A.

Orion A represents the nearest high-mass
star-forming cloud, and as a result, it has been the focus of an intense
observational efforts carried out at multiple
wavelengths and spatial resolutions \citep{gen89}.
The large-scale distribution of CO and its isotopologs has been mapped
repeatedly as radio telescopes improved
in sensitivity and resolution
\citep{kut77,mad86,bal87,cas90,sak94,nag98,wil05,rip13,nis15,kon18}.
Additional molecular species have been mapped by \cite{kau17} using the
Five College Radio Astronomy Observatory (FCRAO),  
\cite{nak19} using the Nobeyama 45 m radio 
telescope, and
\cite{yun21} using the Taeduk Radio Astronomy Observatory (TRAO)
14 m antenna. 
More focused mapping of the so-called integral-shaped filament (ISF), where 
the formation of high-mass stars is taking place, has been done in  C$^{18}$O by \cite{sur19},
H$^{13}$CO$^+$ by \cite{ike07}, N$_2$H$^+$ and HC$_3$N by 
\citet[with further high resolution N$_2$H$^+$ mapping carried out by \citealt{hac17a} 
and  \citealt{hac18}]{tat08},
NH$_3$ by \cite{fri17} as part of the 
Green Bank Ammonia Survey (GAS), and HCN-HNC by \cite{hac20}.

In contrast with Orion A, the California molecular cloud has only recently been
recognized as a distinct star-forming region \citep{lad09}, so its molecular 
emission has received less attention. 
Maps of most of its CO emission have been presented by 
\cite{guo21} and \cite{lew21}, while maps of other tracers have been restricted
to the brightest regions of the cloud, namely 
L1478 (\citealt{chu19}, in CS, N$_2$H$^+$, and HCO$^+$) and
L1482 (\citealt{alv21}, in N$_2$H$^+$, HCO$^+$, and HNC).
Multiline observations of a selection of dense cores selected from {\em Herschel} 
continuum data have been presented by \cite{zha18}.

\section{Observations}

\subsection{Sampling method}

As mentioned in the Introduction and discussed with detail in Paper I, we used the
stratified random sampling technique to select a representative list of cloud 
positions that will be subject to molecular-line observations.
We used the H$_2$ column density as a proxy for the
emission, an approach that was tested in Paper I and is consistent
with the expectation from principal component analysis of different clouds,
which shows that column density is the main predictor of the molecular line intensity 
\citep{ung97,gra17}.
For California and Orion A, \cite{lad17} and \cite{lom14}, respectively, have produced
high-quality H$_2$ column density maps using
far-IR continuum data obtained with the {\em Herschel} Space Observatory \citep{pil10},
and we relied on them for our application of the stratified random sampling.
These maps are complementary to the column density map produced by the same group for Perseus \citep{zar16} and used in Paper I.
All these maps have been ultimately derived from maps 
of dust emission and absorption properties, so the $N$(H$_2$) determination depends on
assumptions about the gas-to-dust ratio and conversion between extinction bands. 
For this, we followed
\cite{lom14}, \cite{zar16}, and \cite{lad17} and assumed the
standard coefficients determined by \cite{boh78}, \cite{sav79}, and \cite{rie85}.

Following Paper I, we divided the range of reliable H$_2$ column densities ($\gtrsim 1.5\times 10^{21}$~cm$^{-2}$, 
\citealt{zar16}) into logarithmically spaced bins of 0.2~dex width.
To reach the maximum column density measured for California and Orion A
($\approx 5\times 10^{22}$~cm$^{-2}$ and $\approx 3\times 10^{23}$~cm$^{-2}$,
respectively), we required 8 and 12 column density bins.
Each of these bins was sampled by choosing 10 random positions, so as a result, 
a total of 80 positions were chosen to
sample the California cloud and 120 positions were chosen for Orion A.
The location of these positions is shown in Fig.~\ref{map_calif} and \ref{map_oria}
superposed on the H$_2$ column density maps of the clouds.
Coordinates of the target positions, together with values of the column
density and the main line intensities are provided in Tables \ref{tbl_master_calif} and \ref{tbl_master_oria}.

\subsection{IRAM 30m telescope observations}
\label{sec_iram_obs}

We observed our target positions in the California and Orion A clouds 
using the IRAM
30 m diameter telescope during three periods in 2018 November, 2019 July,
and 2020 December. The setup was identical to that used in Paper I to sample
Perseus:
the 83.7-115.8 GHz frequency band 
was observed
combining two tunings of the Eight MIxer Receiver (EMIR; \citealt{car12}), which was
 followed by the fast Fourier Transform Spectrometer (FTS;
\citealt{kle12}) to provide a frequency resolution of 200 kHz
($\approx 0.6$~km~s$^{-1}$). For the brighter Orion A cloud, we also 
observed several frequency windows in the range 213.7–267.7 GHz selected for
containing higher-J transitions of species observed at 3mm, such as 
CO(2--1), HCN(3--2), and CS(5--4).
These higher frequency observations were also carried out with the EMIR receiver followed
by the FTS spectrometer at a frequency resolution of 200 kHz 
($\approx 0.25$~km~s$^{-1}$ at the operating frequency).
In addition, selected positions of both California and Orion A 
were observed in HCO$^+$(1--0), HCN(1--0), and C$^{18}$O(2--1) 
with high velocity resolution 
(0.03-0.07~km~s$^{-1}$) using the VESPA autocorrelator to determine 
line shapes and check for self-absorption features.

All survey positions were observed 
in frequency switching mode with throws of $\pm 7.7$~MHz
and total integration times of approximately 10 minutes after combining 
the two linear polarizations of the receiver. Calibration of the atmospheric 
attenuation was carried out observing the standard sequence of 
sky-ambient-cold loads 
every 10-15 min, pointing was corrected every two hours approximately
by making cross scans of bright continuum sources, and focus was corrected using
bright continuum sources at the beginning and several times during the observing session.
The resulting spectra were folded, averaged, and baseline subtracted using 
the CLASS reduction program\footnote{\url{http://www.iram.fr/IRAMFR/GILDAS}}.
The data were also converted into the  
main beam brightness scale using the recommended telescope beam 
efficiencies\footnote{\url{https://publicwiki.iram.es/Iram30mEfficiencies}},
which range from 0.81 at 86~GHz to 0.59 at 230~GHz.
The use of a main beam brightness scale follows standard practice in single-dish calibration, although it may represent an overcorrection
when applied to the emission from the outermost parts of the clouds,
which can be extended over several degrees. 
An alternative calibration choice for this emission would be to 
include the contribution from the telescope error beam, which in the IRAM 30m 
telescope has three components with widths
up to $2,000''$, close to the size of the Moon.
The coupling efficiency of this error beam is about 0.93 at 86~GHz and 0.84 
at 230~GHz \citep{kra13}, which represent an increase with respect to 
the main beam efficiency
of 15 and 42\%, respectively. Using these error beam efficiencies
has therefore little effect on the 3 mm wavelength data, which constitute
the bulk of our survey, although it may affect the 1 mm wavelength 
data if an accurate calibration is required.
Even at 1 mm, however,
the use of error beam efficiencies is probably justified only for
the outer parts of the cloud, and its use could potentially introduce an artificial calibration discontinuity at the transition between the compact and extended 
emission regimes.
For this reason, we preferred to use a single calibration scale based 
on the main beam brightness temperature,
with the caveat that the 1 mm intensities likely have an increased level of 
uncertainty.
In this main beam brightness scale, the typical rms level of the spectra 
is 7-14 mK per $0.6$~km~s$^{-1}$ channel at 100 GHz.

As in Paper I, our analysis of the emission 
relies on the velocity-integrated intensity of the different molecular lines,
which hereafter will be referred to as $I$.
In cases of no detection, we integrated the emission inside the velocity range
at which $^{13}$CO was detected, since this species was detected in all the bins 
and its velocity range always agreed with that of the weaker tracers in case of 
simultaneous detection.
To simplify the velocity integration, we first re-centered all the spectra to zero 
velocity using the centroid of the $^{13}$CO line as a reference, and then we integrated 
the emission in a  common
velocity range. For spectra with multiple hyperfine components, such as HCN(1--0)
and N$_2$H$^+$(1--0), we added together the contribution of all the components to derive a
 single 
intensity value. The uncertainty of the integrated intensity was estimated by propagating 
the contribution of the rms noise in the spectrum over the window of integration, 
although we found that for weak and undetected
lines, the dominant source of uncertainty was the presence of small-level ripples in the 
spectrum baseline, which often exceeded the propagation of the thermal noise by a factor 
of a few.
Additional sources of uncertainty that likely dominate the intensity of the brightest lines
are calibration errors and errors in the telescope efficiencies. Following Paper I, we 
modeled these contributions by adding in quadrature a 10\% error to the 
integrated intensities.
The resulting values for all the lines discussed in this paper are presented 
in Tables \ref{tbl_master_calif} and \ref{tbl_master_oria}.

\section{Results}

\subsection{CO intensity versus H$_2$ column density}

\begin{figure*}
        \centering
    \includegraphics[width=\hsize]{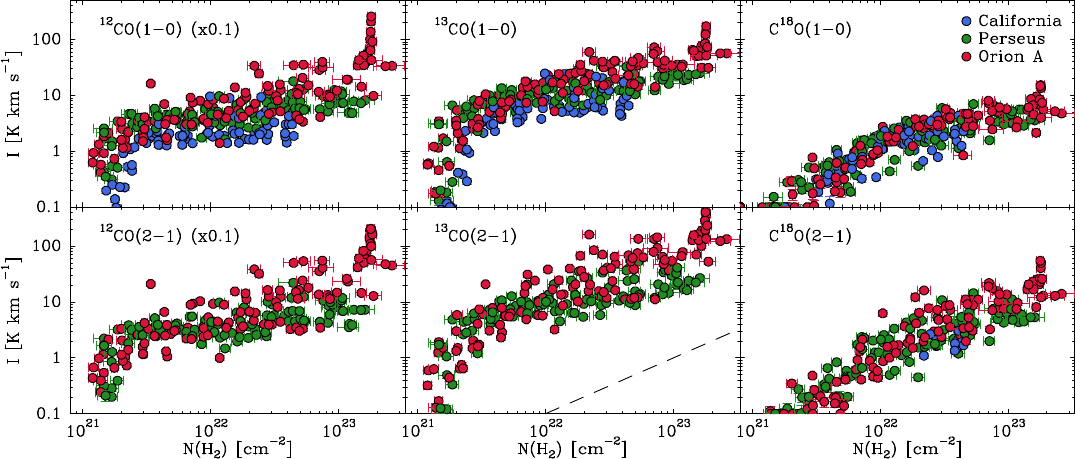}
        \caption{Velocity-integrated intensities of the $J$=1--0 (top) and 
        $J$=2--1 (bottom) lines of the main CO isotopologs
        as a function of H$_2$ column density. 
        The data are color-coded by cloud:
        blue circles for California, green for Perseus, and red for Orion A. 
        The dashed line
        in the $^{13}$CO(2--1) panel indicates the slope of a linear trend. No
        $J$=2--1 data were taken for the California cloud apart from 
        13 C$^{18}$O spectra from high column density positions.
        The lowest value of the intensity scale (0.1~K~km~s$^{-1}$) approximately
        corresponds to the line detection limit.}
        \label{fig_co1}
\end{figure*}

We started our analysis by comparing the emission of the different 
CO isotopologs in our three target clouds.
Figure~\ref{fig_co1} presents velocity-integrated intensities for 
the $J$=1--0 and $J$=2--1 transitions of $^{12}$CO, $^{13}$CO, and 
C$^{18}$O as a function of H$_2$ column density 
(no $J$=2--1 data were taken for the California cloud apart from a small number
of C$^{18}$O spectra).
As the plots show, the intensity of each transition correlates 
significantly with $N$(H$_2$)
over the approximately two orders of magnitude spanned by this parameter
(approximately from 10$^{21}$ to 10$^{23}$ cm$^{-2}$).
Since each observed position was selected randomly among all cloud positions in 
the same
column density bin, the correlation indicates that in each cloud,
the value of $N$(H$_2$) is by itself
a good predictor of the CO line intensity irrespective of the spatial
location of the position.
As discussed in Paper I, our correlations between $^{12}$CO(1--0)
and $^{13}$CO(1--0) intensities with $N$(H$_2$) in Perseus match 
well the correlations between the same parameters
derived from the full maps of \cite{rid06}. For the Orion A cloud,
\cite{yun21} have recently presented intensity correlations
for $^{13}$CO(1--0), C$^{18}$O(1--0), and several dense-gas tracers,
all based on full maps of the cloud. As shown with detail in 
Appendix~\ref{app_yun}, these 
full-map correlations match well the results from our sampling observations,
reinforcing the idea that the
correlations are not an artifact of the sampling technique, 
but a common property of the clouds.

The plots in Fig.~\ref{fig_co1} also show that the correlation
between the different intensities and the H$_2$ column density is similar in the
three clouds,
with the caveat that California spans a narrower range of $N$(H$_2$) than Perseus 
and Orion A.
There are indeed noticeable differences between the clouds, and
they are further discussed below,
but the impression from the panels of Fig.~\ref{fig_co1} is that the main 
trends in the intensity versus $N$(H$_2$) correlation are common to the three clouds. 
In all three clouds, for example, the intensities of the $^{12}$CO and
$^{13}$CO lines drop abruptly below an $N$(H$_2$) of about 
$2 \times 10^{21}$~cm$^{-2}$, which 
corresponds to $A_{\mathrm V}\sim 2$~mag. (A similar drop may occur in 
C$^{18}$O but 
is not noticeable due to the lower signal to noise of the lines.)
This sharp drop likely results from the photodissociation of the
CO molecules by the interstellar radiation field in the cloud outer layers, 
as modeled in Paper I for the case of Perseus. A similar drop
has been observed in other clouds, like Taurus \citep{pin08},
and is predicted by photodissociation models \citep{van88,wol10}.

Interior to the photodissociation boundary, the intensity of the
$^{12}$CO(1--0) and $^{13}$CO(1--0) lines has a 
flatter-than-linear dependence on $N$(H$_2$), an effect that likely
results from the combined high optical depth and thermalization of these lines. 
For the thinner C$^{18}$O lines, the flattening at 
$N(\mathrm{H}_2) > 10^{22}$~cm$^{-2}$ likely results from the freeze-out of the CO molecules at the high densities characteristic of the high 
N(H$_2$), as discussed in detail in Paper I for the case of Perseus.

\subsection{Temperature effects and temperature-corrected CO intensities }
\label{sect_temp}
\label{sect_co_tkcorr}

In addition to similarities, Fig.~\ref{fig_co1} shows systematic 
differences between the distribution of the CO intensity  
in the different clouds. 
The most noticeable one is the enhanced scatter and slightly
steeper slope of the Orion A data at high column densities,
an effect that is especially prominent in the J=2--1
lines of $^{12}$CO and $^{13}$CO.
Since CO is thermalized and therefore
sensitive to temperature, it is tempting to attribute the observed 
differences in the CO emission to differences in the distribution
of the gas temperature across each cloud.
These differences are known to exist in Orion A
(e.g., \citealt{nag98, nis15, fri17}), and they are likely
larger than in California and Perseus due to its
higher star-formation rate \citep{lad10}.
To further investigate the effect of temperature variations in the CO emission,  
we needed to first estimate the gas temperature at each cloud position.
In Paper I we used the C$^{18}$O(2--1)/C$^{18}$O(1--0) ratio to
estimate a relatively constant gas temperature of around 11~K in 
Perseus, and a similar method can be used to estimate the temperature in Orion A.
For California, however, our C$^{18}$O(2--1) observations 
only covered a minority of the cloud positions due to time constraints,
so we could not use the C$^{18}$O line ratio to derive the 
temperature across the cloud.
As an alternative, 
we explored several other methods for determining the gas temperature 
using data available for the three clouds.
These methods include
using the peak intensity of the optically thick $^{12}$CO(1--0) line, using 
the HCN/HNC ratio as recently proposed by \cite{hac20}, and using an empirical
scaled-version of the 
dust temperature that has been calibrated to match the C$^{18}$O line ratio predictions for the positions where it is available.
Appendix~\ref{app_temp_est} describes the details 
of each method an evaluates its quality comparing the results 
with those of the C$^{18}$O(2--1)/C$^{18}$O(1--0) line ratio, which we 
consider the best available temperature indicator.
As can be seen there, the dust temperature method
supplemented with the NH$_3$-derived temperature estimates for
the central part of Orion A from \cite{fri17} provides the 
best results, and for this reason, it is the method of choice for
carrying out the temperature corrections described below.

\begin{figure*}
        \includegraphics[width=\hsize]{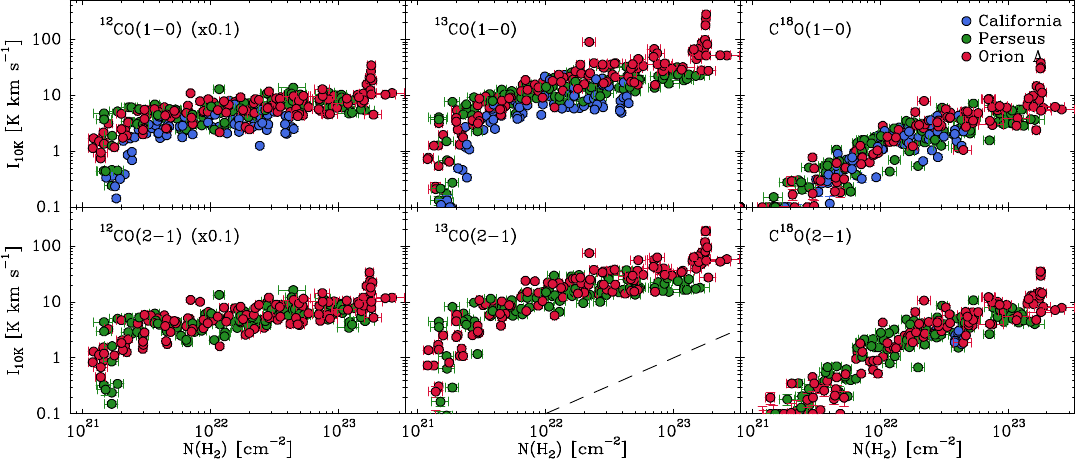}
        \caption{Temperature-corrected intensities of the $J$=1--0 (top) and 
        $J$=2--1 (bottom) lines of the main CO isotopologs
        as a function of H$_2$ column density in
        California, Perseus, and Orion A (blue, green, and red circles, respectively).
        Note the lower level of scatter 
        and the better inter-cloud agreement compared to the
        uncorrected intensities shown in
        Fig.~\ref{fig_co1}. The dashed line
        in the $^{13}$CO(2--1) panel indicates the slope of a linear trend. }
        \label{fig_co_tcorr}
\end{figure*}

Once we had an estimate of the gas temperature at each cloud position,
we used its value to correct the different
line intensities for temperature variations across the cloud.
Calculating a temperature correction factor for each position 
required determining how the intensity of the emerging
lines change as a function of temperature, which is a nontrivial problem
given the complex dependence of the intensity
on multiple physical parameters. After exploring several options, 
we found that the best solution was 
 to use a radiative transfer model,
like the one presented in Paper I, and to predict the dependence  
of the different line intensities on temperature under realistic cloud conditions.
A full description of the method
is presented in Appendix~\ref{app_temp}, while here we summarize its
most relevant aspects.
The Perseus model presented in Paper I 
assumes isothermal gas at 11~K and a simple parameterization of 
the physical and chemical structure of the cloud, and is able to reproduce simultaneously
the emission from the different observed lines.
To use this model to determine how the intensity of the different lines varies
as a function of the gas temperature, we reran it
using a grid of temperatures that ranges from 8~K to 100~K.
Dividing each model intensity by the intensity obtained
at a reference temperature of 10~K,
we derived a series of correction factors $f_{10\mathrm{K}}(T_\mathrm{k})$.
Using these correction factors, we can convert any observed intensity into
the expected intensity that the gas would emit if it were at 10~K. We define this 
``corrected'' intensity as 
\begin{equation}
I_{10\mathrm{K}} \equiv \frac{I}{f_{10\mathrm{K}}(T_\mathrm{k})},
\end{equation}
where $I$ is the observed intensity and 
$f_{10\mathrm{K}}(T_\mathrm{k})$ is the model correction factor.
Values for these factors 
for each transition observed in our survey are presented 
in Tables~\ref{tbl_tcorr_dense} and \ref{tbl_tcorr_co}.

Using the above factors, we converted the intensities of the CO isotopologs
into their expected values for gas at 10~K, and present the results 
as a function of $N$(H$_2$) in Fig.~\ref{fig_co_tcorr}.
Compared to the uncorrected intensities, the Orion A
corrected intensities show a significant decrease of dispersion, which is a 
factor of 1.6 for the main isotopologs. 
As a result, the rms
dispersion of the Orion A CO intensities is typically 0.2 dex,
which is similar to the rms estimated for Perseus and California.
These two clouds have also been temperature corrected, but the effect of 
the correction is minimal due to the clouds almost-constant temperature.

In addition to decreasing the dispersion, the temperature correction 
helps equalize the line intensities of the three clouds.
For all transitions, the temperature-corrected 
intensities
of Perseus and Orion A are practically indistinguishable, apart from the spike of 
points near $2 \times 10^{23}$~cm$^{-2}$ in Orion A caused by the wings of the 
Orion Kleinmann–Low (Orion-KL)
outflow. The line intensities of the California cloud are also close
to those of Perseus and Orion A, although they cluster at 
the lower end of the range spanned by the other two clouds. 
This equalization of the temperature-corrected intensities
strongly suggests that most differences seen between the uncorrected 
intensities are due to differences in the gas temperature,
and that
once the temperature has been equalized (even using an approximate
method like ours) a common pattern of emission emerges from the data.
The existence of this common emission pattern 
indicates that after H$_2$ column density, the
gas kinetic temperature is the next physical parameter that controls the intensity 
of the CO lines,
and that when it is adjusted, the intensity of the CO isotopolog lines
lines can be predicted
for each $N$(H$_2$) value with a precision close to 0.2 dex.

The only peculiar 
emission feature in Fig.~\ref{fig_co_tcorr} that remains unaffected  
by the temperature correction 
is the sudden increase in the intensity of the Orion A lines 
at column densities close to $2 \times 10^{23}$~cm$^{-2}$.
As mentioned before, this intensity increase is associated with the Orion-KL
molecular outflow, and is accompanied by the appearance of prominent
high velocity wings in the CO spectra. For the optically 
thick CO lines, the appearance of the wings
increases the velocity range available for the CO emission to escape, 
and this effect contributes to the large increase seen in intensity.
The intensity increase, however, can also be seen in the optically thin
C$^{18}$O emission, so it must be accompanied by a local increase 
in the CO abundance that
likely results from the action of outflow shocks and
dust heating caused by the feedback of the massive stars in Orion-KL.
The very extended N$_2$H$^+$ emission seen toward the ISF 
\citep{hac18} indicates the presence of large-scale CO freeze-out, 
and this process is likely being 
reversed in the vicinity of Orion-KL
by the action of the stellar feedback. 
While smaller abundance enhancements occur toward individual
low mass stars (such as the L1448 region in Perseus), the extreme 
effect of the Orion-KL outflow is unique to the Orion A cloud, and 
seems to represent a different regime of chemical abundance driven by
high-mass star formation. 
This regime coincides with the small fraction of gas having 
column 
densities higher than of $2\; 10^{23}$~cm$^{-2}$,
which corresponds to a mass density of 0.94~g~cm$^{-2}$, a value
practically equal to the 1~g~cm$^{-2}$ threshold for star formation
proposed by \cite{kru08}. This different chemical regime is therefore likely
to occur in other regions of high-mass formation, although further observations
are required to reach a firm conclusion.

\subsection{Quantifying the intensity comparison}

So far, our comparison of the intensity profiles 
in the different clouds has been purely qualitative.
To quantify it, we need a statistical tool that tests
whether two distributions of points are equal.
A convenient choice for this is the test proposed by \citet[the FF test hereafter]{fas87}, which generalizes the classical 
Kolmogorov-Smirnov test to multidimensional distributions of data. 
As the Kolmogorov-Smirnov test, the FF test determines
the probability (p-value) that any two 2D samples
arise from the same underlying distribution.
To apply the test, it
is customary to choose a threshold probability $\alpha$ 
(typically 0.05) so that if the p-value is lower than $\alpha$,
the hypothesis that the two samples arise from
the same distribution (the null hypothesis) can be considered as rejected. 
As often stressed in the literature (e.g., \citealt{pre92}), 
p-values larger than $\alpha$ do not guarantee that the two samples arise from
exactly the same distribution, but that they are not different enough to reach
a definitive conclusion. Given this caveat, and the arbitrary choice of the
threshold value $\alpha$, our use of the FF test does not pretend
to prove or disprove mathematically that any two intensity distributions 
are completely equivalent,
but to explore how significant any differences may be.

\begin{table*}
\caption[]{FF test p-values for a comparison between the emission
of the different CO isotopologs in each pair of clouds.\tablefootmark{a}
\label{tbl_co_fftest}}
\centering
\begin{tabular}{lcccccc}
\hline\hline
\noalign{\smallskip}
\multicolumn{7}{c}{No temperature correction} \\
\noalign{\smallskip}
\hline
\noalign{\smallskip}
Clouds & CO(1--0) & $^{13}$CO(1--0) & C$^{18}$O(1--0) & 
CO(2--1) & $^{13}$CO(2--1) & C$^{18}$O(2--1) \\
\noalign{\smallskip}
\hline
\noalign{\smallskip}
California\tablefootmark{b} - Perseus & $4.9\; 10^{-9}$ & $4.8\; 10^{-6}$ &  \bm{$5.2\; 10^{-1}$} & --
 & -- &  -- \\
California - Orion A & $7.8\; 10^{-10}$ & $2.9\; 10^{-9}$ & \bm{$9.4\; 10^{-2}$} &
--  & -- & -- \\
Perseus - Orion A & $2.8\; 10^{-2}$ & $1.7\; 10^{-2}$ & \bm{$2.2\ 10^{-1}$} &
$4.7\; 10^{-3}$ & $3.9\; 10^{-3}$ & \bm{$8.6\; 10^{-2}$} \\
\noalign{\smallskip}
\hline
\noalign{\smallskip}
\multicolumn{7}{c}{With temperature correction} \\
\noalign{\smallskip}
\hline
\noalign{\smallskip}
Clouds & CO(1--0) & $^{13}$CO(1--0) & C$^{18}$O(1--0) & 
CO(2--1) & $^{13}$CO(2--1) & C$^{18}$O(2--1) \\
\noalign{\smallskip}
\hline
\noalign{\smallskip}
California - Perseus & $2.5\; 10^{-6}$ & $5.0\; 10^{-6}$ & \bm{$5.2\; 10^{-1}$} & 
 -- & -- & -- \\
California - Orion A & $8.4\; 10^{-9}$ & $2.0\; 10^{-5}$ & \bm{$9.7\; 10^{-2}$} &
 -- & -- & -- \\
Perseus - Orion A & \bm{$4.9\; 10^{-1}$} & $2.1\; 10^{-2}$ & \bm{$3.1\ 10^{-1}$} &
\bm{$6.7\; 10^{-1}$} & $4.4\; 10^{-2}$ & \bm{$3.0\; 10^{-1}$} \\
\hline
\noalign{\smallskip}
\end{tabular}
\tablefoot{
\tablefoottext{a}{Boldface values indicate a probability higher than the
0.05 threshold and are consistent with indistinguishable distributions.}
\tablefoottext{b}{No $J$=2--1 data available for the California cloud.}
}
\end{table*}

To apply the FF test to our data, 
we used its implementation in the {\tt R} statistical
program\footnote{\url{https://www.R-project.org/}} 
\citep{rco18} as presented by \cite{pur21}, which
provides a fast and straightforward evaluation of the test p-value.
Since the FF test compares distributions in pairs, we applied the 
test to each pair of clouds and to each CO transition observed 
in both clouds. To avoid any bias caused by the different 
$N$(H$_2$) extent of the
different clouds, we restricted the comparison between clouds
to the intensity values inside a common range of $N$(H$_2$), which for
comparisons involving the California cloud is
$2\times 10^{21}$-$4.8\times 10^{22}$~cm$^{-2}$, and for comparisons between 
Perseus and Orion A is 
$2\times 10^{21}$-$1.5\times 10^{23}$~cm$^{-2}$. 
Finally, to study the effect of the 
temperature correction, we 
ran the FF test before and after applying the correction.

Table~\ref{tbl_co_fftest} summarizes the p-values derived from the FF test
for each cloud pair and transition for which data are available.
Values in bold face indicate
probabilities that exceed the standard 0.05 threshold value and are therefore
consistent with the two intensity distributions being equivalent.
As can be seen, 
each pair comparison involving a C$^{18}$O transition exceeds the 0.05 threshold
irrespectively of whether the temperature correction has been applied or not,
although the temperature-corrected p-values are slightly larger
and therefore suggest a better match.
This low sensitivity of the FF test to the temperature correction 
results from the small value of the correction for the optically thin
C$^{18}$O lines (Fig.~\ref{fig_tcorr}), and has the advantage of 
making the C$^{18}$O  line comparison almost independent 
of the temperature estimate.
Since the C$^{18}$O emission is optically thin, the similarity between 
the intensity distributions of
the three clouds suggests that the clouds share a similar CO
abundance distribution.

Another conclusion that can be derived from the data in Table~\ref{tbl_co_fftest} 
is that,
in contrast with the C$^{18}$O results, the comparison of the
CO and $^{13}$CO emission from California with both Perseus and Orion A
returns p-values significantly lower than the 0.05 threshold irrespectively
of the use of the temperature correction. If we interpret the 
C$^{18}$O comparison
as indicative that the three clouds have similar CO abundance profiles, the 
low p-values of the
CO and $^{13}$CO comparison suggest that either the excitation or 
the optical depth in California is different from that in Perseus and Orion A.
Figure~\ref{fig_co_tcorr} suggests that the CO and $^{13}$CO 
intensities in California are lower than in
Perseus and Orion A, and an experiment of multiplying the California
intensities by a factor of 1.5 brings the clouds in better agreement.
This result suggests that we are either overcorrecting the California intensities
or that the California intensities are intrinsically lower due to optical depth 
effects, such as self-absorptions caused by the narrower lines.
Further observations of these isotopologs are needed to reach a firm conclusion.

Finally, the p-values in Table~\ref{tbl_co_fftest} confirm the success
of the temperature 
correction equalizing the intensity of the main CO isotopolog 
in Perseus and Orion A.
Before applying a temperature correction, the FF test for 
both CO(1--0) and CO(2--1) returns p-values below the 0.05 threshold,
while the p-values after applying the temperature correction increase to
 0.49 and 0.67 for $J$=1--0 and 2--1, respectively.
This large change in the p-value reflects the large size of the 
temperature correction for the very optically thick lines of the
main isotopolog, and illustrates the need of
considering temperature variations when comparing the CO emission from
different clouds. 

Somewhat surprisingly, the temperature correction
does not bring the p-value of the $^{13}$CO intensities over the 0.05
threshold, although it increases the p-value of the 2--1 transition by one
order of magnitude. The reason for this failure may represent an intrinsic
difference between the clouds, possibly caused by different isotopic 
fractionation \citep{lan80,ish19}, or simply represent a failure of our
temperature correction caused by an underestimate of the $^{13}$CO
optical depth.

To summarize, our analysis shows that the FF test is a useful tool 
for quantitatively comparing the distribution of intensities between clouds.
When applied to the C$^{18}$O emission, the FF test shows that the
emission from the three clouds is 
statistically indistinguishable, and since this emission is optically thin, 
it likely indicates that the clouds have similar CO abundance distributions.
The observed differences between the distributions of the optically thick
$^{12}$CO and $^{13}$CO emissions likely arise from differences in the internal
temperature structure of the clouds.

\subsection{Traditional dense-gas tracers}
\label{sect_trad_dense}

\begin{figure*}
    \centering
        \includegraphics[width=0.7\hsize]{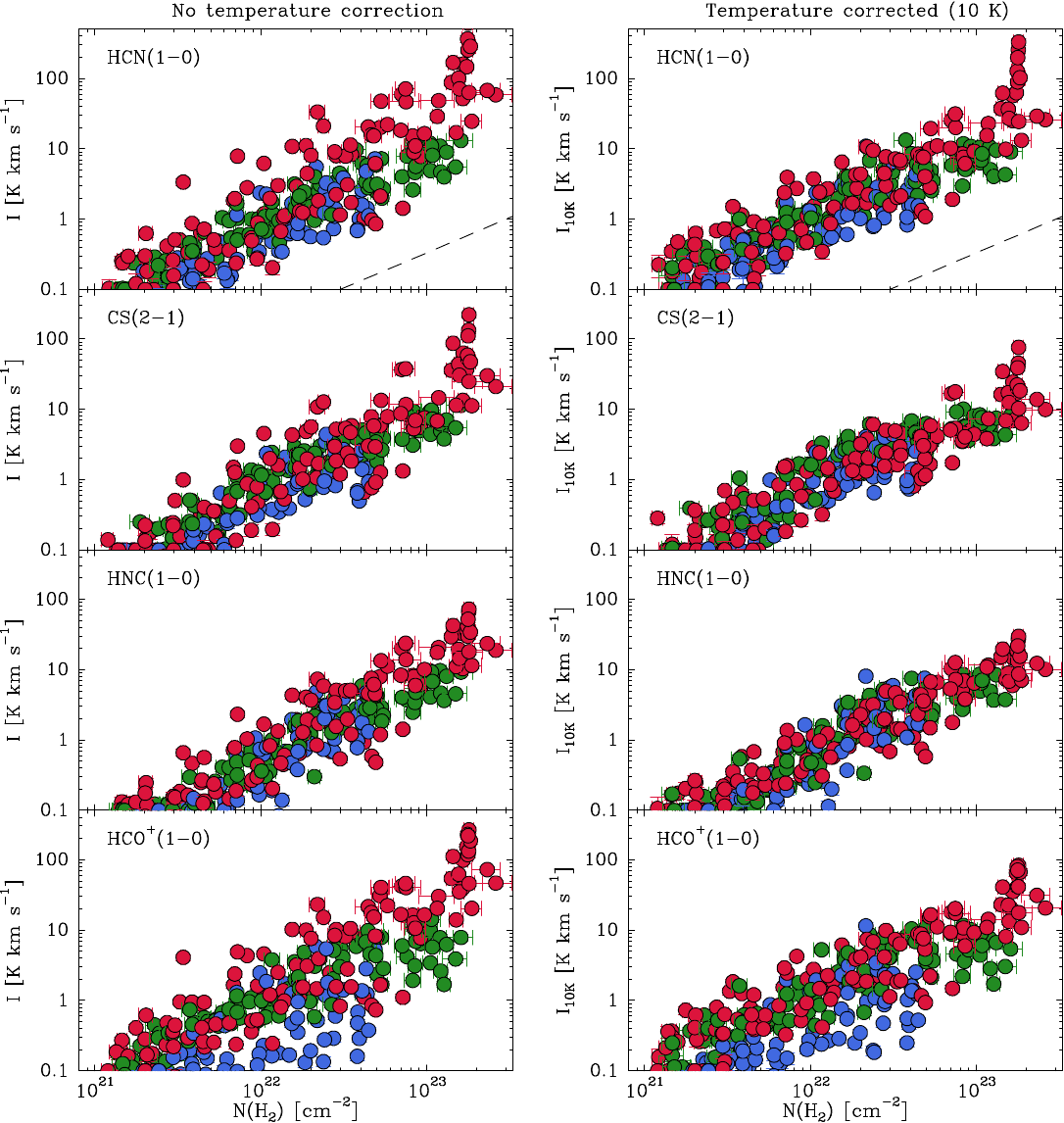}
        \caption{Velocity-integrated intensities of the traditional dense-gas tracers HCN(1--0), CS(2--1), HNC(1--0), and HCO$^+$(1--0)
        as a function of H$_2$ column density. {\em Left panels:} Original
        uncorrected data. {\em Right panels:} Data after applying the
        correction factors described in Appendix~\ref{app_temp} to 
        simulate emission at a constant temperature of 
        10~K. Note the decrease in dispersion and the better agreement between clouds
        after the temperature correction. The data are color-coded as
        in previous figures: blue for California, green for Perseus, and red for 
        Orion A. The dashed line in the top panels represents a linear trend
        for comparison.} 
        \label{fig_dens}
\end{figure*}

We now turn our attention to several molecular species that combine 
a high
dipole moment with bright 3 mm wavelength lines: HCN, CS, HCO$^+$, and HNC. 
Following Paper I, we collectively refer
to these species as ``traditional dense-gas tracers'' since they
have been used in the past as indicators of high-density gas in
molecular clouds (e.g., \citealt{eva99}).
Recent research shows that these tracers are less selective of dense
material than initially thought  
\citep{kau17,pet17,wat17,shi17,eva20,taf21,dam23},
although they are still widely used 
by the extragalactic community due to their bright emission lines
\citep{gao04,use15,gal18,jim19}

Since the traditional dense-gas tracers combine a large abundance and a
high dipole moment, their low-$J$ lines are expected to be optically
thick over a significant fraction of any cloud. For our selected lines, 
this expectation is confirmed by the value of
the intensity ratio between the main and
a rare isotopolog (H$^{13}$CN, C$^{34}$S, H$^{13}$CO$^+$, and HN$^{13}$C). 
As shown in  Fig.~\ref{fig_dense_ratios}, all these ratios
systematically lie below the expected abundance ratio by a large margin,
indicating that the
main isotopolog lines must be highly saturated over most of the cloud positions.

As it happened with CO, the high optical depth of the traditional dense-gas tracer lines makes their emission potentially sensitive to temperature 
variations. To evaluate this effect, we present
in Fig.~\ref{fig_dens} plots of the integrated intensity of 
HCN(1--0), CS(2--1), HNC(1--0), and HCO$^+$(1--0) as a function of 
H$_2$ column density for two cases: uncorrected data (left panels) and data
corrected for temperature variations using the factors described in 
Appendix~\ref{app_temp} (right panels).
As can be seen, the temperature-corrected data present a lower degree of
scatter and a better agreement between the emission from the three clouds 
compared to the uncorrected data. We interpreted this equalizing effect of the 
temperature correction as an indication
that most of the inter and intra cloud differences in the emission 
seen in the uncorrected data
arise from variations in the gas kinetic temperature.
An exception to this equalizing effect are the large increases in the HCN(1--0) and CS(2--1) 
intensity at column densities around $2 \times 10^{23}$~cm$^{-2}$ in
the Orion A cloud. 
As we will see the next section using the more abundance-sensitive rare isotopologs,
these intensity increases are likely the result 
of abundance enhancements caused by high-mass star-formation feedback.

Since our goal was to investigate how the molecular emission is 
generated
inside the clouds,
we focused on the temperature-corrected data and searched for additional
features in the intensity profiles.
A first feature to notice is the tight correlation 
between the intensity of all traditional dense-gas tracers and the 
H$_2$ column density.
The correlation is stronger in
HCN(1--0), CS(2--1), and HNC(1--0), and is characterized by 
values of the Pearson's $r$ coefficient in the range of 0.87-0.90 
(Table~\ref{tbl_dense_slopes}).
The HCO$^+$(1--0) intensities, on the other hand, 
present a significantly higher degree of
scatter, which is mostly caused by the California data lying significantly
below the Orion A and Perseus data. The Pearson's $r$ coefficient 
is 0.80, which still indicates a strong level of correlation between
the HCO$^+$ emission and the H$_2$ column density.
Paper I already noted that the HCO$^+$(1--0) intensities in Perseus presented 
a higher degree of scatter than the other lines, and the new 
California data show an even larger dispersion. As shown in the next section,
the intensity of the H$^{13}$CO$^+$ rare isotopolog presents little scatter 
and a tight correlation with $N$(H$_2$), suggesting that the larger scatter of 
the main isotopolog lines
results from optical depth effects. An inspection of high velocity resolution 
HCO$^+$(1--0) spectra taken toward selected positions reveals the presence of significant
self-absorption features in some spectra that artificially truncate the 
emission and therefore lower the intensity.

\begin{table}
\caption[]{Statistics of the (logarithmic) 
correlation between the temperature-corrected 
intensity of traditional dense-gas tracers and H$_2$ column density.
\label{tbl_dense_slopes}}
\centering
\begin{tabular}{lcc}
\hline\hline
\noalign{\smallskip}
Transition & $r$-Pearson & Slope  \\
\noalign{\smallskip}
\hline
\noalign{\smallskip}
HCN(1--0)  & 0.87 & $1.12\pm 0.04$  \\
CS(2--1) & 0.88 & $1.07\pm 0.03$  \\
HCO$^+$(1--0) & 0.80 & $0.97\pm 0.04$ \\
HNC(1--0) & 0.90 & $1.16\pm 0.03$ \\ 
\noalign{\smallskip}
\hline
\noalign{\smallskip}
\end{tabular}
\end{table}

\begin{table}
\caption[]{FF p-values for traditional dense-gas tracers.
\label{tbl_dense_fftest}}
\centering
\begin{tabular}{lcccc}
\hline\hline
\noalign{\smallskip}
\multicolumn{5}{c}{No temperature correction} \\
\noalign{\smallskip}
\hline
\noalign{\smallskip}
Clouds & HCN(1-0) & CS(2-1) & HNC(1-0) & HCO$^+$(1-0) \\
\noalign{\smallskip}
\hline
\noalign{\smallskip}
Cal-Pers & $2.3\; 10^{-2}$ & $3.7\; 10^{-2}$ & \bm{$7.2\; 10^{-2}$} &  
$1.0\; 10^{-6}$  \\
Cal-Ori& $4.1\; 10^{-2}$ & \bm{$1.1\; 10^{-1}$} & $1.7\; 10^{-2}$ &  
$2.2\; 10^{-4}$  \\
Pers-Ori & $3.0\; 10^{-2}$ & \bm{$3.5\; 10^{-1}$} & 
\bm{$2.5\; 10^{-1}$} &  $3.0\; 10^{-2}$  \\
\noalign{\smallskip}
\hline\noalign{\smallskip}
\multicolumn{5}{c}{With temperature correction} \\
\noalign{\smallskip}
\hline
\noalign{\smallskip}
Clouds & HCN(1-0) & CS(2-1) & HNC(1-0) & HCO$^+$(1-0) \\
\noalign{\smallskip}
\hline
\noalign{\smallskip}
Cal-Pers & \bm{$1.4\; 10^{-1}$} & $2.8\; 10^{-2}$ & \bm{$1.7\; 10^{-2}$} &  
\bm{$7.2\; 10^{-5}$}  \\
Cal-Ori & $3.7\; 10^{-2}$ & \bm{$2.7\; 10^{-1}$} & $1.9\; 10^{-2}$ &  
$1.9\; 10^{-5}$  \\
Pers-Ori & \bm{$2.8\; 10^{-1}$} & \bm{$5.9\; 10^{-2}$} & 
\bm{$5.1\; 10^{-1}$} &  \bm{$3.3\; 10^{-1}$}  \\
\noalign{\smallskip}
\hline
\end{tabular}
\tablefoot{
Boldfaced p-values exceed the 0.05 null-hypothesis threshold.
}
\end{table}

In addition to a strong correlation with $N$(H$_2$), Fig.~\ref{fig_dens} shows 
that the temperature-corrected intensity of the
traditional dense-gas tracers has an 
approximately linear dependence on column density (indicated by the dashed 
lines in the top panels). 
The only significant deviation from this trend
occurs near $N$(H$_2$) = $2\times 10^{23}$~cm$^{-2}$, where several tracers
present a sudden intensity increase coincident with the
position of the Orion Nebula Cluster (ONC) and the 
Orion-KL outflow. The origin of this increase seems to be a 
combination of abundance variations in some species and a drop in the optical
depth due to the wide outflow wings caused by the outflow 
(as seen in the isotopic ratios of Fig~\ref{fig_dense_ratios}). In the 
next section we use the intensity distribution of the 
rare isotopologs to disentangle these two contributions. 
For the current analysis, we focused on the slope of the distribution 
as determined from a least squares fit to the combined data of the three clouds 
(including the anomalously bright positions at high column densities).
The fit results, summarized in the third column of
Table~\ref{tbl_dense_slopes}, show that the slopes lie in a narrow range 
(0.97-1.16) and are therefore very close to unity, as expected from the 
inspection of Fig.~\ref{fig_dens}. 
This quasi-linear slope can be followed without significant changes 
until the emission reaches the detection limit
($\approx 0.1$~K km s$^{-1}$),
indicating the absence of breaks until the H$_2$ column density reaches
its lowest values ($\approx 10^{21}$~cm$^{-2}$).
The radiative transfer model presented in Paper I showed that the gas volume
density approximately follows the column density (Sect. 5.1), so the 
continuity of the slope in $N$(H$_2$) indicates that there is no particular density 
at which the  emission of the traditional dense-gas tracers suddenly 
changes behavior or disappears. 
Because of this, we can say
that the tracers are sensitive to the gas density (since their intensity
gradually increases with this parameter), but that they are not selective
of any density value, since the emission depends continuously on this
parameter.
In addition, 
molecular clouds present probability distribution functions of column
density that increase nonlinearly toward low column densities \citep{lom15}. 
As a result, the cloud-integrated intensity of any
traditional high-density tracer is expected to be dominated by the contribution of
the cloud low-density regions, as previously shown by different authors
\citep{kau17,pet17,wat17,shi17,eva20,taf21,jon23}.

To finish our analysis, we again used  the FF test
to quantify the similarities between the emission distributions
in the three clouds. 
Table~\ref{tbl_dense_fftest} presents the p-values derived using the FF test
for both cases of no temperature correction and temperature correction.
In agreement with the expectation from Fig.~\ref{fig_dens}, using 
temperature correction improves the agreement between the clouds,
reinforcing the idea that despite its simplicity, 
the correction partially
compensates for differences in the gas temperature between the clouds.
The better agreement of the temperature-corrected intensities also suggests that
there are important similarities between the emission of the three clouds. The
largest differences occur again when comparing California with Perseus and Orion A,
especially for the case of HCO$^+$(1--0). For the other lines, the FF test 
produces a mix of p-values slightly larger or smaller than the 0.05 threshold,
confirming that 
there are noticeable similarities between the emission of the clouds, but not necessarily
to the point of making them indistinguishable.
In contrast with the peculiar behavior of the California cloud, a comparison 
between the intensity-corrected intensities in 
Perseus and Orion A produces
p-values that are always
larger than the 0.05 threshold for the four 
observed lines (although only marginally for CS(2--1)).
The emission of Perseus and Orion A therefore appears indistinguishable
when comparing positions in their common range of H$_2$ column density. 

\begin{figure*}
    \centering
        \includegraphics[width=\hsize]{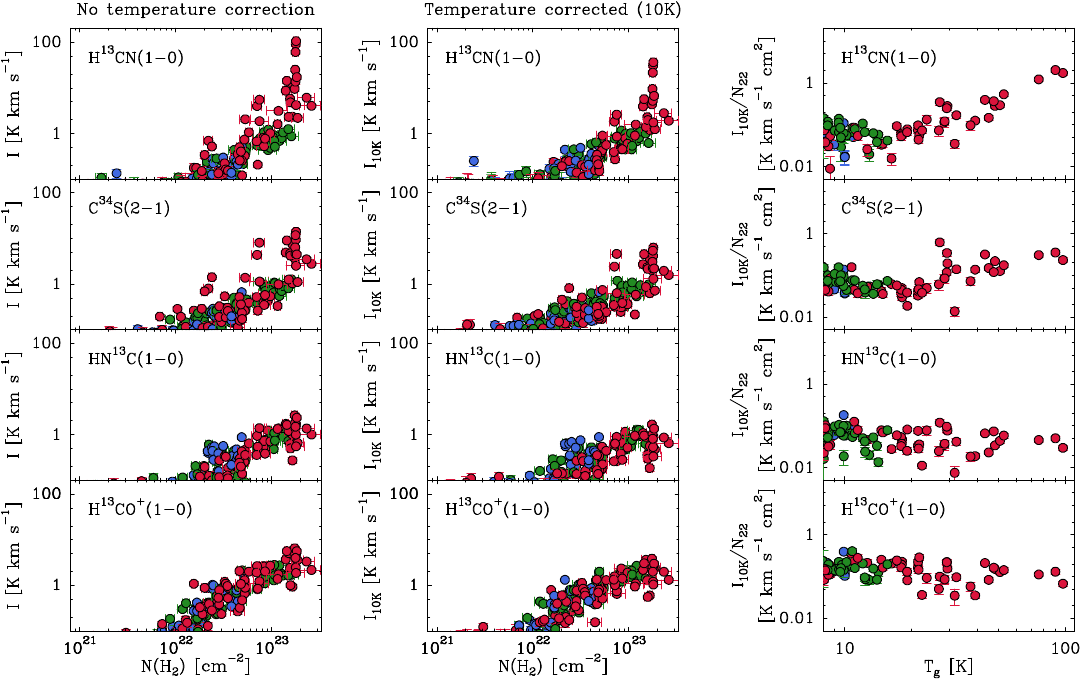}
        \caption{Intensity distributions of the rare isotopologs of the traditional
        dense-gas tracers represented in Fig.~\ref{fig_dens}.
        {\em Left panels:} Original uncorrected data.
        {\em Middle panels:} Data after applying the
        correction factors described in Appendix~\ref{app_temp} to 
        simulate emission at a constant temperature of 10~K.
        {\em Right panels:} Ratio between the temperature-corrected intensity and the
        H$_2$ column density (in units of $10^{22}$~cm$^{-2}$) as a function of gas
        temperature. All data are color-coded as in previous figures.}
        \label{fig_dens_isotop}
\end{figure*}

\subsection{Rare isotopologs of the traditional dense-gas tracers}
\label{sec_rare}

The lines of the traditional dense-gas tracers discussed in the previous section
are optically thick and therefore
relatively insensitive to possible abundance variations. 
To investigate these variations, we had to rely on less
abundant isotopologs, such as H$^{13}$CN, C$^{34}$S, HN$^{13}$C, and 
H$^{13}$CO$^+$, whose lines are
likely to be optically thin as suggested by
the relative intensities of the hyperfine components of H$^{13}$CN(1--0).

Figure~\ref{fig_dens_isotop} presents the intensity distribution
of the most abundant rare isotopologs of the traditional dense-gas tracers studied in the previous section. 
The left and middle panels show their
emission as a function of H$_2$ column 
density both using  uncorrected intensities
(left panels) and intensities corrected for temperature variations
using the prescriptions detailed in Appendix~\ref{app_temp} (middle panels).

As can be seen, the rare-isotopolog lines are weaker than the
main-species lines by about one order of magnitude, and as a result,
their detection in our survey
is limited to H$_2$ column densities larger than 
approximately $10^{22}$~cm$^{-2}$.
In this range of detection, the intensity of most species 
is very similar in the three clouds.
For H$^{13}$CO$^+$(1--0), this is in contrast with the behavior of
the main isotopolog, which as we saw in Fig.~\ref{fig_dens},
presents strong excursions toward low intensities
in California.
The lack of similar excursions in the H$^{13}$CO$^+$(1--0) intensity
supports the interpretation
that HCO$^+$(1--0) suffers 
from optical depth effects, most likely from the
strong self-absorptions known to affect the narrower California lines.

If we compare the left and middle panels of Fig.~\ref{fig_dens_isotop}, we 
notice that applying 
the temperature correction decreases some of the
intensity excursions seen in H$^{13}$CN(1--0) and C$^{34}$S(2--1),
but has otherwise little effect on the data. This is due to the small
value of the correction, which is
typically less than a factor of 2 (Fig.~\ref{fig_tcorr}),
and as a result, it cannot remove the strong intensity
increase seen in H$^{13}$CN(1--0) and C$^{34}$S(2--1)
toward  $N(\mathrm{H}_2) \approx 2\times 10^{23}$~cm$^{-2}$.

To investigate the intensity increase of
H$^{13}$CN(1--0) and C$^{34}$S(2--1) in Orion A, the
right panels of Fig.~\ref{fig_dens_isotop}  present
plots of the ratio between the temperature-corrected
intensity and the H$_2$ column density
as a function of the gas kinetic temperature
for the four rare isotopologs.
Positions with gas temperature below 20~K are present in all three
clouds, while all warmer positions are located in the ISF of Orion A, especially in 
the vicinity of Orion KL and the ONC.
As can be seen, the ratios for HN$^{13}$C and H$^{13}$CO$^+$ 
(bottom panels) remain approximately constant, with a possible slight 
decrease at intermediate temperatures, 
despite a factor of 10 variation in the gas temperature.
On the other hand, the ratios for H$^{13}$CN and C$^{34}$S (top panels), 
present significant correlations with the gas temperature 
for values larger than approximately 20~K. In the 20-100~K temperature range, 
the ratio for H$^{13}$CN increases by about one order of
magnitude while the ratio for C$^{34}$S increases about a factor of 5.

Since the correlations seen in the right panels
of Fig.~\ref{fig_dens_isotop} involve
intensities already corrected for temperature variations, their 
most likely origin must be differences in the abundance of the species
as a function of the gas temperature.
If this is case, the approximately constant intensity/column density ratios of 
H$^{13}$CO$^+$ and HN$^{13}$C suggest that these two species
are relatively immune to temperature-related abundance variations, 
which makes them stable tracers of the column density.
The increase in the intensity/column density ratios of
H$^{13}$CN and C$^{34}$S, on the other hand, strongly indicates that the abundance
of these two species depends sensitively on temperature once this parameter 
exceeds a threshold value of around 20~K. The larger increase of 
the H$^{13}$CN intensity/column density ratio indicates that this species is
more sensitive to temperature than C$^{34}$S, and
that its abundance is expected to vary as a result of
gas temperature increases caused by the action of star formation.

Our finding of an HCN abundance enhancement at high temperatures is in good
agreement with previous research on the chemistry of Orion A, which has found a
significant increase of the HCN/HNC ratio with the gas kinetic
temperature \citep{gol81,sch92,hac20}.
Although not fully understood, this 
increase likely results from the activation of temperature-sensitive 
neutral-neutral reactions that alter the total abundance of the two species and
their abundance ratio \citep{her00,gra14}.
Less work has been carried out on the possible temperature dependence 
of the abundance of CS.
We note however that most positions with high CS abundance in Orion A 
present evidence for high-velocity wings caused by the Orion-KL outflow, and that
chemical surveys of bipolar outflow gas often show significant abundance
enhancements of both HCN and CS (but little or no enhancement of 
HNC and HCO$^+$; see \citealt{bac97}, \citealt{taf10}, and \citealt{lef21}).
Clearly more work is needed to understand the 
different contributions to the abundance behavior of HCN, CS, HNC, and HCO$^+$
at high temperatures. For the purposes of our study, our main conclusion is 
that the temperature enhancement resulting from star-formation feedback
introduces a new chemical regime in the cloud gas 
that seems to coincide with the onset of high-mass star formation.

\begin{table*}
\caption[]{FF p-values for isotopologs of 
traditional dense-gas tracers.
\label{tbl_isot_dense_fftest}}
\centering
\begin{tabular}{lcccc}
\hline\hline
\noalign{\smallskip}
\multicolumn{5}{c}{No temperature correction} \\
\noalign{\smallskip}
\hline
\noalign{\smallskip}
Clouds & H$^{13}$CN(1-0) & C$^{34}$S(2-1) & HN$^{13}$C(1-0) & 
H$^{13}$CO$^+$(1-0) \\
\noalign{\smallskip}
\hline
\noalign{\smallskip}
Cal-Pers & \bm{$6.3\; 10^{-1}$} & \bm{$4.1\; 10^{-1}$} & \bm{$1.5\; 10^{-1}$} &  
\bm{$8.3\; 10^{-1}$}  \\
Cal-Ori & \bm{$5.3\; 10^{-1}$} & \bm{$4.0\; 10^{-1}$} & $4.3\; 10^{-2}$ &  
\bm{$8.0\; 10^{-1}$}  \\
Pers-Ori & \bm{$3.3\; 10^{-1}$} & \bm{$4.6\; 10^{-1}$} & \bm{$6.6\; 10^{-1}$} &  
\bm{$3.3\; 10^{-1}$}  \\
\noalign{\smallskip}
\hline
\noalign{\smallskip}
\multicolumn{5}{c}{With temperature correction} \\
\noalign{\smallskip}
\hline
\noalign{\smallskip}
Clouds & H$^{13}$CN(1-0) & C$^{34}$S(2-1) & HN$^{13}$C(1-0) & 
H$^{13}$CO$^+$(1-0) \\
\noalign{\smallskip}
\hline
\noalign{\smallskip}
Cal-Pers & \bm{$3.1\; 10^{-1}$} & \bm{$4.5\; 10^{-1}$} & \bm{$1.2\; 10^{-1}$} &  
\bm{$4.8\; 10^{-1}$}  \\
Cal-Ori & \bm{$1.1\; 10^{-1}$} & \bm{$7.3\; 10^{-1}$} & \ $2.7\; 10^{-3}$ &  
\bm{$3.6\; 10^{-1}$}  \\
Pers-Ori & \bm{$2.2\; 10^{-1}$} &  $4.3\; 10^{-2}$ & \bm{$4.1\; 10^{-1}$} &  
\bm{$3.4\; 10^{-1}$}  \\
\noalign{\smallskip}
\hline
\end{tabular}
\tablefoot{
Boldfaced p-values exceed the 0.05 null-hypothesis threshold.
}
\end{table*}

To conclude our analysis of the rare isotopologs, we present 
in Table~\ref{tbl_isot_dense_fftest} the results of the
FF test for all combinations of the lines and pairs of clouds, 
both before and after applying the temperature correction. Since
the isotopologs are not detected at H$_2$ column densities lower than
about $10^{22}$~cm$^{-2}$, only values larger than this threshold
have been considered. As in previous FF tests, the California data
have been compared with Perseus and Orion A data having column densities 
smaller or equal to $4.8\times 10^{22}$~cm$^{-2}$, and the 
Perseus data have been compared with Orion A data up to a column density of
$1.5\times 10^{23}$~cm$^{-2}$.
As can be seen in the table, all p-values for the uncorrected comparison
exceed the 0.05 threshold except for the comparison of 
HN$^{13}$C(1--0) between California and Orion A, which returns a
value of 0.04.
After applying the temperature correction, most p-values remain larger
than the 0.05 threshold, except for the already mentioned HN$^{13}$C(1--0)
and the comparison of C$^{34}$S(2--1) between Perseus and Orion, whose
p-value drops by one order of magnitude with respect to the uncorrected
comparison. This drop seems anti-intuitive in view of the plots of 
Fig.~\ref{fig_dens_isotop}, and seems to occur because the temperature correction
decreases the dispersion of the data, so any slight difference between the
emission of the clouds becomes more significant. Still, the fact that the
majority of p-values exceed the 0.05 threshold is an indication that
while small differences may exist, the emission from the three clouds
presents strong similarities when similar column densities are compared.
The main differences between the clouds seem therefore to arise from the 
fact that they reach very different peak column densities.

\subsection{N$_2$H$^+$ and the onset of molecular freeze-out}
\label{sec_n2hp}

\begin{figure*}
    \centering
        \includegraphics[width=\hsize]{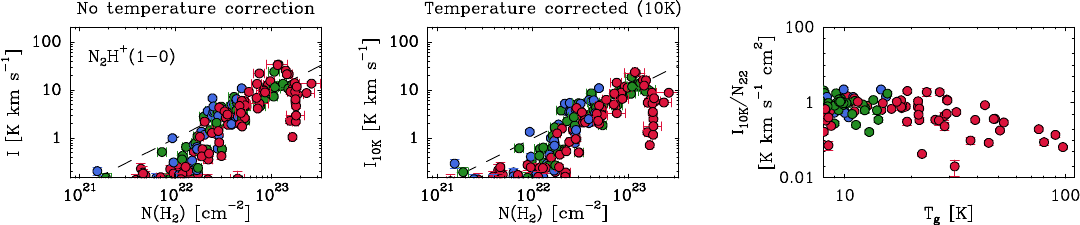}
        \caption{Distributions of N$_2$H$^+$(1--0) integrated intensity.
        {\em Left panel:} Original uncorrected data.
        {\em Middle panel:} Data after applying the
        correction factors described in Appendix~\ref{app_temp} to 
        simulate emission at a constant temperature of 10~K.
        {\em Right panel:} Ratio between the temperature-corrected intensity and the
        H$_2$ column density (in units of $10^{22}$~cm$^{-2}$) as a function of gas
        temperature. All data are color-coded as in previous figures.}
        \label{fig_n2hp}
\end{figure*}

In contrast with the traditional dense-gas tracers, 
which freeze out onto the cold dust grains at
high densities and low temperatures, N$_2$H$^+$ 
remains in the gas phase and its abundance is enhanced
at high densities, likely as a result of the
freezing out of CO 
\citep{kui96,cas99,aik01,taf02,lee04}. Since N$_2$H$^+$ also has a high 
dipole moment \citep{hav90}, it has become a tracer of choice for identifying
the dense and cold condensations responsible for star formation in clouds 
\citep{ber07}.
To illustrate the behavior of the N$_2$H$^+$ emission in the three clouds of our sample,
we present its distribution in Fig.~\ref{fig_n2hp}. As in previous figures,
the left and middle panels show the
distribution of intensity as a function of H$_2$ column density
both without temperature correction and after applying the temperature
correction factors described in Appendix~\ref{app_temp}.
Since the 
temperature correction for N$_2$H$^+$ never exceeds a factor of 2 
(Fig.~\ref{fig_tcorr}), the two distributions in the figure 
look very similar.

As can be seen from  Fig.~\ref{fig_n2hp},
the distribution of N$_2$H$^+$(1--0) intensity differs significantly 
from that of the traditional dense-gas tracers. 
It follows an almost linear correlation with $N$(H$_2$)
at high column densities, but
below $2\times 10^{22}$~cm$^{-2}$
it drops nonlinearly with column density, and remains 
undetected at values below $10^{22}$~cm$^{-2}$.
This sudden change in the emission is unique to  N$_2$H$^+$(1--0),
and makes this species a highly selective tracer of the dense gas in a cloud.

\begin{table}
\caption[]{FF test p-values for N$_2$H$^+$(1--0).
\label{tbl_n2hp_fftest}}
\centering
\begin{tabular}{lcccc}
\hline\hline
\noalign{\smallskip}
Clouds & No $T_\mathrm{k}$ corr. &  With $T_\mathrm{k}$ corr.\\
\noalign{\smallskip}
\hline
\noalign{\smallskip}
Cal-Pers & \bm{$8.8\; 10^{-1}$} & \bm{$6.1\; 10^{-1}$} \\
Cal-Ori & \bm{$3.7\; 10^{-1}$} & \bm{$2.0\; 10^{-1}$} \\
Pers-Ori & \bm{$3.1\; 10^{-1}$} & \bm{$2.1\; 10^{-1}$} \\
\noalign{\smallskip}
\hline
\end{tabular}
\tablefoot{
Boldfaced p-values exceed the 0.05 null-hypothesis threshold.
Only data with $N$(H$_2$)>$10^{22}$~cm$^{-2}$ used.
}
\end{table}

As Fig.~\ref{fig_n2hp} shows, the distribution of N$_2$H$^+$(1--0) 
emission in the three clouds is very similar independently of whether a
temperature correction has been applied or not.
This good agreement is confirmed by the FF test results reported in 
Table~\ref{tbl_n2hp_fftest}, which show that the p-value in all comparisons 
exceeds the 0.05 threshold both with and without temperature correction.
The good match between the clouds in the nonlinear region between 
$10^{22}$ and $2\times 10^{22}$~cm$^{-2}$ 
is especially remarkable because the nonlinear change is
likely caused by the onset of CO freeze-out, a process that is 
also  sensitive to volume density because the freeze-out time depends on
the collision time between molecules and dust grains \citep{leg83}. 
The similar location of the N$_2$H$^+$(1--0) change  
in the three clouds suggests that the critical density for CO freeze-out is
reached at a similar column density in all of them despite their
very different peak H$_2$ column densities and star-formation rates.

Also noticeable in the distribution of N$_2$H$^+$(1--0)
is the sharp drop toward the highest column densities reached in Orion A
at about $10^{23}$~cm$^{-2}$ coinciding with the ONC and Orion BN/KL.
This drop has been previously noticed by a number of authors, including
\cite{tat08}, \cite{kau17}, \cite{hac18}, and \cite{yun21},
and occurs at the same column densities at which 
HCN and CS present their abundance increase 
associated with the high-temperature gas.
To investigate the effect of temperature in the  N$_2$H$^+$ abundance, we
again calculated the ratio between the N$_2$H$^+$ intensity and the 
H$_2$ column density, 
and present the result as a function of the gas temperature in
the right panel of Fig.~\ref{fig_n2hp}. To make this plot, 
we restricted the comparison to H$_2$
column densities larger than 
$2\; 10^{22}$~cm$^{-2}$ since this is the approximate range at which
the N$_2$H$^+$ intensity depends quasi-linearly with $N$(H$_2$). 
As can be seen, 
the intensity-column density gradually decreases in gas hotter than 
about 30~K, and by 100~K, it has decreased by about one order of 
magnitude with respect to its low-temperature value.

Although the N$_2$H$^+$ drop presents more scatter than the increases
in H$^{13}$CN and C$^{34}$S, the trend points again to a change in the
gas chemical composition triggered by the feedback from star formation.
A drop in the N$_2$H$^+$ abundance is indeed expected as a result of the
release of CO from the dust grains due to protostellar heating, 
and has been previously observed toward individual star-forming regions
\citep{joe04,cas12,joe20}.
Observations of additional high-mass star-forming regions are necessary to
confirm this interpretation.

\subsection{Line luminosity estimates from sampling observations and 
comparison with mapping results}
\label{sec_lum}

 So far we have only used the sampling data to study the
distribution of
line intensities as a function of H$_2$ column density.
This type of distribution represents the most immediate output from the sampling 
observations, and as we have seen, provides
a detailed description of the emission properties from a cloud.
The sampling data can also be used to estimate other emission properties, such as
the line luminosity of a cloud, 
which corresponds to the integral of the line intensity over the
cloud surface area. 
This luminosity can be calculated from the sampling data
by adding the contribution from
each column density bin to the product of 
the mean line intensity ($I_n$) times the surface area 
subtended by that bin ($A_n$). In other words, 
\begin{equation}
L = \sum_{n=1}^m I_n\; A_n,
\label{eq_lum}
\end{equation} 
where in our case $n$ runs 
from 1 to $m$= 8, 10, and 12 for California, Perseus, 
and Orion A, respectively.

In practice, the mean line intensity $I_n$ can be estimated by 
averaging all the
spectra observed toward a given column
density bin (ten positions in our survey)
and integrating the emission over the full velocity range.
The surface area $A_n$ subtended by the bin can be estimated 
from the available extinction maps \citep{lom14, zar16, lad17}
by counting the number of pixels
belonging to the bin and multiplying the result
by the pixel area assuming an appropriate cloud distance.
Combining these two quantities,
it is straightforward to use the sampling data to 
estimate the luminosity of any line emitted by a cloud. 

To test whether the above method provides accurate estimates of the
line luminosities, we searched the literature for line
luminosity determinations based on the standard mapping technique
in any of our three target clouds, and we
calculated equivalent luminosity estimates using our sampling data. 
As expected, 
most available luminosity determinations involve CO transitions since 
this molecule has been the tracer of choice for large-scale mapping.

The most complete set of luminosity determinations
that can be used to compare with our sampling estimates 
is that of \cite{lew22}, who have determined 
 $^{12}$CO(1--0) luminosities for California,
Perseus, and Orion A 
as part of their study of 12 nearby molecular clouds.
These authors have used for their estimates
data from the Milky Way survey of \cite{dam01},
and while they
do not explicitly provide the resulting luminosities, those can be trivially
derived from the $\alpha_{\mathrm{CO}}$
conversion factors and the cloud masses presented in their Table~1.
Due to the
large scale of the maps used by \cite{lew22} (see their Figs.~12 and 13), 
the resulting luminosities should be compared with sampling estimates
using the full extent of the clouds.

Additional CO luminosities for the Orion A cloud have been presented 
by \cite{nis15}, who have estimated values for both
the  $J$=1--0 and 2--1 transitions of $^{12}$CO, $^{13}$CO, and  
C$^{18}$O. These luminosities, however, have been 
estimated after applying to the data a noise-reduction mask
that assigns ``zero values at the emission free pixels''
(their Sect.~2.1), and as a result, this set of luminosity
estimates neglects the contribution from the outer parts of the cloud,
which according to our sampling estimates is significant. 
To properly compare luminosities estimated using sampling with
the results from \cite{nis15}, it is therefore 
critical to exclude the fraction of
the cloud that these authors have masked out.
This information is only available for
the $J$=2--1 lines, whose (masked) maps are publicly available
(Sect. 6 in \citealt{nis15}), so our comparison with the sampling
method has to be restricted to the $J$=2--1 transitions. For them, we 
downloaded the maps from the repository, verified the luminosity
values given by \citealt{nis15} in their Table~1, and estimated the 
amount of surface area in each of our 
column density bins that was left unmasked.
Using these values, we used Eq.~\ref{eq_lum} to
derive sampling-based luminosities that can be properly
compared with the values presented by \cite{nis15}.

\begin{figure}
        \centering
        \includegraphics[width=\hsize]{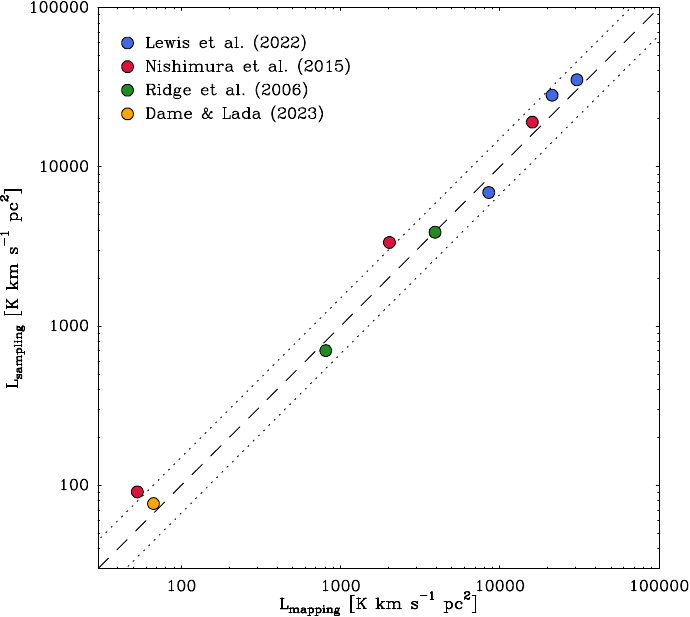}
        \caption{Comparison between line-luminosity estimates 
        based on mapping data from the literature and equivalent 
        estimates using 
        sampling observations. The blue circles represent CO(1--0) luminosities of
        California, Perseus, and Orion A 
        from \cite{lew22}, the red circles the CO(2--1), $^{13}$CO(2--1),
        and C$^{18}$O(2--1) luminosities of Orion A from from \cite{nis15},
        the green circles
        the CO(1--0) and $^{13}$CO(1--0) luminosities of Perseus from
         \cite{rid06}, and the yellow circle the HCN(1--0) luminosity of Perseus
         from \cite{dam23}. The dashed line 
         marks the locus of equal estimates, and the dotted lines correspond to 
         differences of 50\%. All estimates assume the cloud distances given in 
         Sect.~\ref{sec_intro}.}
        \label{fig_luminosity_test}
\end{figure}

A final set of CO luminosities based on mapping observations 
can be derived from the publicly available maps 
of the Coordinated Molecular Probe Line Extinction
and Thermal Emission (COMPLETE) survey of Perseus presented by \citealt{rid06}
These $^{12}$CO(1--0) and $^{13}$CO(1--0)
maps contain more than $10^5$ pixels each, and we 
spatially integrated them to estimate cloud luminosities.
Since the COMPLETE survey did not cover the full extent of the
Perseus cloud
(see the coverage in Fig.~2 of \citealt{rid06}), we 
estimated the amount of area of each bin covered by the COMPLETE 
maps and used these values
to calculate equivalent sampling-based estimates.

Apart from CO, the only species whose line luminosity has been 
estimated in any of our three target clouds is HCN.
\cite{dam23} have recently presented an estimate of the 
HCN(1--0) luminosity from Perseus using a map
made with the 
Center for Astrophysics (CfA) 1.2 m telescope 
that attempts to cover the full extent of the cloud emission.
While this map only extends to a column density equivalent to
our second bin (based on its CO cutoff), these authors have
used their larger CO map to estimate the contribution from the remaining
"weak, unobserved HCN," so we used this corrected luminosity to compare 
with out sampling estimate for the full cloud.

Fig.~\ref{fig_luminosity_test} summarizes the comparison between
the mapping and sampling luminosity estimates
by representing one quantity against the other
(numerical values are given in Table~\ref{tbl_luminosities_comparison}).
The diagonal dashed line indicates the locus of equal luminosity
estimates, and
the parallel dotted lines delimit the region where the mapping and sampling
estimates agree at the 50\% level.
As can be seen, the luminosity estimates 
span almost three orders of magnitude
and systematically cluster along the equal-value dashed
line, indicating 
an overall good agreement between the two methods used to estimate luminosities.

As the figure indicates, the level of agreement between the mapping and sampling 
estimates seems to slightly vary between the 
different data sets, although there is no evidence for significant variations
with the choice of cloud or tracer.
The \cite{lew22} estimates 
(blue circles), which are the only ones that include simultaneously
our three target clouds,
present differences with the mapping results that 
are only at the level of 30\% or less. This is despite the use of of very
different telescopes: the beam solid angle of the CfA 1.2m telescope  
is 625 times larger than that of the IRAM 30m telescope
used for our sampling observations. 

A slightly worse level of agreement is seen in the
comparison between the Orion A sampling results 
and the estimates of \cite{nis15}, which are
represented in the figure by three red circles (corresponding by 
decreasing order to the $J$=2--1 transitions of
$^{12}$CO,  $^{13}$CO, and C$^{18}$O).
The differences between the two data sets are at the 50\% level, 
which is the largest value in all our comparisons.
While we can only speculate as to why the sampling luminosities are
significantly larger than the mapping ones in this case, we 
note that this comparison is the only one involving
1 mm wavelength data. As
mentioned in Sect.~\ref{sec_iram_obs}, our use of the main beam brightness scale
at 1 mm may over calibrate the IRAM 30 m data by about 40\% in 
the case of very extended emission.
The data from \cite{nis15}, on the other hand, were taken with 
the Osaka 1.85m telescope, which has a low sidelobe level,
and whose $T_{\mathrm R}^*$ scale is more appropriate for extended emission
\citep{oni13,nis15}. A difference in the
calibration scheme used to reduce the two data sets
may therefore be responsible for part of the disagreement between the 
estimated luminosities.

A better level of agreement is seen in the comparison with the Perseus data
of \cite{rid06} (green circles), where the differences between the mapping and
sampling luminosities are of 15\% or less. This good agreement is consistent 
with the results from the comparison
between the distribution of line intensities as a function
of $N$(H$_2$) for this data set and our Perseus observations
carried out in Paper I.
A similar level of agreement is seen for the HCN(1--0) luminosity
estimate presented by \cite{dam23} for Perseus
(orange circle), suggesting that the ability of the sampling method
to estimate line luminosities is not limited to observations
of the CO transitions.

In addition to showing that sampling observations can provide accurate
estimates of the line luminosities, the comparison with the mapping data
shows that our choice of column density bins
captures the bulk of the emission even in the case of the very extended CO lines. 
This is supported by the good match with the luminosities from \cite{lew22}.
Since these authors extracted their maps from a Milky Way survey,
we can safely assume that their maps were not artificially limited 
by mapping coverage, but by the natural extent of the clouds. 
Our sampling luminosities
agree with those of \cite{lew22} to better than 30\%, so
any emission coming from regions outside the lowest column density bin 
of our sampling must contribute negligibly to the total cloud output.
This result was expected from the finding of sharp drops in the CO intensity
toward the lowest column density bins  of all the clouds,
which were interpreted as resulting from molecular photodissociation
caused by the external UV radiation field. It suggests that any 
molecular gas outside the lowest bin in our sampling is likely to be 
CO dark.

\begin{figure*}
        \centering
        \includegraphics[width=\hsize]{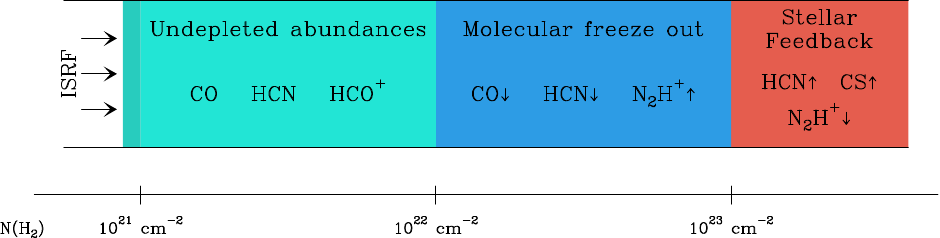}
        \caption{Cartoon view of the main chemical regimes identified in 
        the observed clouds. The labels refer to
        representative species observed in each regime, and the arrows indicate
        their abundance trends. The numerical values
        on the horizontal scale represent approximate estimates of the column density
        at which the transition between the regimes occurs. {This plot is inspired by Fig. 12
        of \cite{ber07}.}
        }
        \label{plot_cloud_model}
\end{figure*}

To summarize, our comparison shows that the stratified random 
sampling method can be used to estimate line luminosities
that agree with previously published values at the a level 
typically better than about 30\%.
This result is reassuring  in view of the large differences 
in the observing techniques, spatial coverage, calibration schemes,
and size of the telescopes involved in the comparison.
Our comparison also shows that to obtain accurate luminosity estimates,  
both techniques require special care. 
The sampling technique requires using
high-quality extinction maps to estimate
the surface area subtended by each column density bin, 
and sampling the emission down to column densities of around 
1-$2 \times 10^{21}$~cm$^{-2}$.
The mapping technique requires good spatial coverage of the cloud
and the ability to account for the weak emission from the outer 
parts of the cloud, whose contribution to the luminosity 
is not negligible due to their large surface area
and cannot be masked out.

Having validated the method, we calculated luminosities for all the 
lines studied in the previous sections, and the results are summarized in 
Table~\ref{tbl_luminosities}. It should be noted that the N$_2$H$^+$(1--0)
values represent only lower limits because the emission of this line 
was not detected in the
outer layers of the cloud, and their potential contribution cannot be estimated with 
our data.
Further discussion
of the HCN(1--0) luminosities and their relation with the amount of
dense gas in the clouds is deferred to Sect.~\ref{sec_alpha_hcn}.

\begin{table}
\caption[]{Line luminosity estimates.
\label{tbl_luminosities}}
\centering
\begin{tabular}{lc|lc}
\hline\hline
\noalign{\smallskip}
Line & Luminosity & Line & Luminosity \\ 
 & (K km s$^{-1}$ pc$^2$) & & (K km s$^{-1}$ pc$^2$) \\
\noalign{\smallskip}
\hline
\noalign{\smallskip}
\multicolumn{4}{c}{California} \\
\noalign{\smallskip}
\hline
\noalign{\smallskip}
$^{12}$CO(1--0) & 35,200 & CS(2--1) & 78 \\
$^{13}$CO(1--0) & 4,220 & HNC(1--0) & 68 \\
C$^{18}$O(1--0) & 226 & HCO$^+$(1--0) & 173 \\
HCN(1-0) & 210 & N$_2$H$^+$(1--0) & $\ge 8$ \\
\noalign{\smallskip}
\hline
\noalign{\smallskip}
\multicolumn{4}{c}{Perseus} \\
\noalign{\smallskip}
\hline
\noalign{\smallskip}
$^{12}$CO(1--0) & 6,900 & $^{12}$CO(2--1) & 5,850 \\
$^{13}$CO(1--0) & 1,160 & $^{13}$CO(2--1) & 938 \\
C$^{18}$O(1--0) & 83 & C$^{18}$O(2--1) & 61 \\
HCN(1--0) & 77 & HNC(1--0) & 30 \\
CS(2--1) & 61 & HCO$^+$(1--0) & 82 \\
N$_2$H$^+$(1--0) & $\ge 9$ & & \\
\noalign{\smallskip}
\hline
\noalign{\smallskip}
\multicolumn{4}{c}{Orion A} \\
\noalign{\smallskip}
\hline
\noalign{\smallskip}
$^{12}$CO(1--0) & 28,200 & $^{12}$CO(2--1) & 20,100 \\
$^{13}$CO(1--0) & 4,070 & $^{13}$CO(2--1) & 4,150 \\
C$^{18}$O(1--0) & 234 & C$^{18}$O(2--1) & 316 \\
HCN(1--0) & 524 & HNC(1--0) & 210 \\
CS(2--1) & 236 & HCO$^+$(1--0) & 544 \\
N$_2$H$^+$(1--0) & $\ge 36$ & & \\
\noalign{\smallskip}
\hline
\end{tabular}
\end{table}

\section{Discussion}

\subsection{The three main chemical regimes of a molecular cloud}

The similar dependence on $N$(H$_2$) and $T_{\mathrm{gas}}$ of the line
intensities in
California, Perseus, and Orion A suggests that the three clouds share a similar
chemical structure, and that this structure can be described using
$N$(H$_2$) and $T_{\mathrm{gas}}$ as the main physical parameters.
In this section we combine the results of our 
analysis of the intensity distributions in the three clouds 
with the results of the radiative transfer model of the Perseus cloud 
presented in Paper I to 
determine the main characteristics of the chemical structure of the clouds.
A cartoon view of the proposed structure is presented in 
Fig.~\ref{plot_cloud_model}.

We start our discussion with the cloud outermost layers.
Their chemical composition can only be studied
using the few species that are bright enough to be detected toward the 
lowest column density bins. As shown in Figs.~\ref{fig_co1} and
 \ref{fig_co_tcorr}, 
the emission of both $^{12}$CO and $^{13}$CO is detected at all column densities, 
and presents a sharp change around $N$(H$_2$) = 
1-$2\times 10^{21}$~cm$^{-2}$, which is equivalent to a visual extinction
of $A_{\mathrm V} = 1$-2~mag . Similar sharp changes 
of the CO emission have been
seen toward the edges of other molecular clouds \citep{pin08,rip13}, and they
most likely result from the photodissociation of CO 
by the interstellar radiation field \citep{van88,wol10}.

As discussed in Paper I, the Perseus data also show hints that the intensity
of some traditional dense-gas tracers present an outer change
similar to CO, and that
some UV-sensitive species, such as C$_2$H and CN present slight
outer abundance enhancements in agreement with the expectations from
models of photodissociation regions  \citep{cua15}. 
All these effects indicate that in the three clouds,
the column density value of 1-$2 \times 10^{21}$~cm$^{-2}$
marks the approximate boundary between the outer UV-dominated regime and the
shielded cloud interior, where most molecular species seem to keep approximately
constant abundances (as suggested by the radiative transfer model of
Paper I). In Fig.~\ref{plot_cloud_model}, we 
represent this region as the outermost layer of the cloud, and 
label its interior as the regime of "undepleted abundances."

The next significant change in the cloud chemical composition seems to
occur after the column density has increased by about one order of magnitude.
The plots of N$_2$H$^+$ intensity show that this tracer experiments an order of
magnitude increase between $10^{22}$ and $2\times 10^{22}$~cm$^{-2}$, after which
the intensity approximately follows quasi-linearly $N$(H$_2$) (Fig.~\ref{fig_n2hp}).
As mentioned in Sect.~\ref{sec_n2hp}, this sharp increase in the N$_2$H$^+$ 
abundance is expected to correspond to the onset of CO freeze-out onto the 
dust grains, and is a consequence
of the gradual increase in the gas volume density as the
column density increases. The occurrence of this onset at similar
column densities in California, Perseus, and Orion A points to a
similar increase in the gas volume density as a function of column density
in the three clouds, which suggests that the clouds share an important 
similarity in their internal structure. Since the freeze-out of CO is accompanied 
by a similar
freeze-out of other carbon species such as CS, HCO$^+$, and HCN 
\citep{kui96,taf06}, we interpreted
the column density value of $10^{22}$~cm$^{-2}$ as an approximate boundary
of the second regime of the cloud, which we call the molecular freeze-out regime (Fig.~\ref{plot_cloud_model}).

The final chemical regime suggested by our observations has a less sharp boundary,
but approximately corresponds to column densities in excess of $10^{23}$~cm$^{-2}$.
This regime is only present in the Orion A cloud since California and Perseus
do not reach such high values of $N$(H$_2$), and is represented by the  
high-mass star-forming regions in the ISF.
As discussed in Sect.~\ref{sec_rare}, these regions
present elevated gas temperatures (>30~K) and abundance enhancements in
selected species such as HCN and CS, and are likely the result of 
high-temperature chemistry triggered by high-mass star formation.
This regime therefore represents the effect of stellar feedback on the
cloud gas, and its properties likely depend less systematically 
on $N$(H$_2$) than the other two
regimes due to the more stochastic nature of the star-formation
activity. Given that our sampling of column densities larger than 
$10^{23}$~cm$^{-2}$ is limited to only 
Orion A, observations of other high-mass star-forming regions are
still required to properly characterize this chemical regime.

\subsection{HCN as a dense-gas tracer}
\label{sec_alpha_hcn}

\begin{figure*}
        \centering
        \includegraphics[width=\hsize]{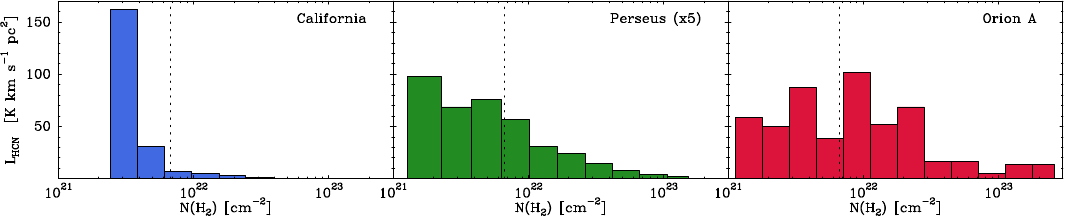}
        \caption{Contribution to the HCN(1--0) luminosity of the different H$_2$ 
        column density bins to which each cloud has been assigned. The vertical dashed
        line indicates the column density corresponding to $A_{\mathrm K} = 0.8$~mag,
        proposed by \cite{lad10} as the boundary of the cloud dense gas. For the California
        cloud, no HCN(1--0) emission was detected toward the lowest column density bin, so
        its contribution to the luminosity has been set to zero (see the main text).}
        \label{fig_hcn_lum}
\end{figure*}

Due to its bright lines, HCN has become the tracer of choice to estimate the amount
of dense gas in extragalactic studies of star formation. 
In their classical analysis of the HCN emission from a wide variety
of galaxies, \cite{gao04} found a close-to-linear correlation
between the far-IR luminosity and 
the HCN(1--0) luminosity,
and interpreted it as indicating that the star-formation rate of a galaxy depends
on the amount of dense gas traced by HCN. 
To determine the mass of the dense gas
associated with the HCN emission, 
\cite{gao04} used a combination of large velocity gradient (LVG)
radiative transfer and virial analysis, and concluded that
\begin{equation}
M_{\mathrm{dense}} = \alpha(\mathrm{HCN}) \; L_{\mathrm{HCN}},
\label{eq_alpha}
\end{equation}
with a conversion factor $\alpha$(HCN) of around
10~M$_\odot$~(K km s$^{-1}$ pc$^2$)$^{-1}$. Although \cite{gao04}
recognized the approximate nature of their $\alpha$(HCN)
estimate, and stated that an accurate determination required 
``further extensive studies,'' their proposed
value has become a de facto standard
for extragalactic studies of star formation 
(e.g., \citealt{use15,gal18,jim19}).

As mentioned in Sect.~\ref{sect_trad_dense}, recent studies 
of the emission from galactic clouds have shown
than HCN(1--0) is not a truly selective tracer of 
the dense gas since 
its cloud-scale emission is dominated by the contribution
from extended and relatively low density gas 
\citep{kau17,pet17,wat17,shi17,eva20,taf21,dam23}.
While this result calls into question a literal 
interpretation of the $\alpha$(HCN) derivation by
\cite{gao04}, the existence of a 
tight linear correlation between the HCN luminosity and the
far-IR luminosity, a reliable tracer of the star formation rate, still
indicates that HCN traces either the amount of dense gas or a gas property
that is closely connected to the star-forming material. 
Understanding the origin of the \cite{gao04} relation therefore remains
an open question whose answer requires investigating the 
origin of the HCN(1--0) emission from local clouds.

To investigate the role of the HCN emission as a dense-gas tracer,
we used 
our sampling data to evaluate the HCN conversion factor
in the California, Perseus, and Orion A clouds.
Before discussing the results, it should be noted that there are
multiple definitions of the HCN conversion factor in
the literature (e.g., see Table A.2. in \citealt{shi17}), so it is important to 
first clarify how the conversion factor is defined. 
Broadly speaking, two types of
definitions have been proposed depending on whether the factor is considered as
a global cloud parameter or a a local quantity that varies with the line of sight.
Each type of definition focuses on a different aspect of the relation between the 
HCN emission and the cloud gas, and provides a useful clue to the origin 
of the HCN emission. We therefore discuss them in sequence.

\subsubsection{The global $\alpha_{08}$(HCN) factor}

The global $\alpha$(HCN) factor relates the cloud-integrated HCN(1--0) luminosity 
to the total mass of the dense gas, and follows the spirit of the original 
\cite{gao04}
definition as given in Eq.~\ref{eq_alpha}. This factor is most relevant for 
extragalactic observations since they do not resolve the emission from individual
clouds and therefore need to rely on cloud-integrated quantities.
In their original derivation, \cite{gao04} assumed that the HCN emission was
truly selective of the dense gas, and estimated that $\alpha$(HCN) was
approximately equal to
$2.1\, \langle n(\mathrm{H}_2)\rangle^{1/2}/T_\mathrm{b}$~$M_\sun$(K 
km s$^{-1}$ pc$^2$)$^{-1}$, 
where $\langle n(\mathrm{H}_2)\rangle$ is the average gas density and 
$T_\mathrm{b}$ is the line brightness temperature. 
Assuming that these parameters take values of 
$3\times 10^4$~cm$^{-3}$ and 35~K, respectively, \cite{gao04} derived
the often-used result that 
$\alpha\mathrm{(HCN)} \approx  10$~$M_\sun$(K km s$^{-1}$ pc$^2$)$^{-1}$.
Since we now know that the HCN emission is not truly 
selective of the dense gas \citep{kau17,pet17,wat17,shi17,eva20,taf21,dam23}, 
it has become customary to determine the value of the global 
$\alpha$(HCN) factor by defining
the amount of dense gas using an independent criterion, such as
the amount of mass over
an extinction threshold of $A_\mathrm{K}=0.8$~mag,
which seems to correlate with the star-formation rate of a cloud 
(\citealt{lad10}; see also \citealt{eva14}). From now on, we refer 
to this definition of the conversion factor as $\alpha_{0.8}$(HCN).

In Sect.~\ref{sec_lum} we show how the sampling data can be used to derive
line luminosities, and present estimates for HCN and other species in 
each of out sample clouds (Table~\ref{tbl_luminosities}).
To better understand how the HCN(1--0) luminosity arises from the different 
layers of each cloud, we now present in Fig.~\ref{fig_hcn_lum}
histograms of the luminosity as a function of $N$(H$_2$)
for California, Perseus, and Orion A.
Each bin in the histogram corresponds to a column density bin
of our sampling, 
and the dotted vertical lines mark the  $A_\mathrm{K}=0.8$~mag threshold
used to define the dense gas
($\approx 6.7\times 10^{21}$~cm$^{-2}$; e.g., \citealt{lom14}).
In the California cloud, no HCN(1--0) emission was detected
in the average spectrum of the lowest column density bin, so
we set its line contribution to zero, while in  Perseus and
Orion A the HCN(1--0) emission was detected even in the lowest column
density bin.

As can be seen in the figure, the highest column density bins contribute the least
to the HCN luminosity in each cloud. This occurs because despite their 
brighter lines, these high column density bins cover a very small area,
so their contribution to the total luminosity 
cannot compete with that of the weaker but much more extended emission from the low
column density gas. Using the $A_\mathrm{K}=0.8$~mag threshold
as a boundary for the dense gas, our data indicate that the high density material
only contributes to the total HCN(1--0) luminosity by 8\% in
California, 37\% in Perseus, and 55\% in Orion A (Table~\ref{tbl_alpha_glbl}). 
Our estimate for Perseus 
is very close to the 40\% estimated by \citealt{dam23}
from their mapping observations, and a
similar range of values has been determined for clouds in the inner
and outer Galaxy by \cite{eva20} and \cite{pat22}, respectively.

\begin{table*}
\caption[]{Dense gas masses, HCN(1--0) luminosities, and global HCN conversion factors\tablefootmark{a}.
\label{tbl_alpha_glbl}}
\centering
\begin{tabular}{lccccc}
\hline\hline
\noalign{\smallskip}
Cloud & $M_{08}$ & $L_\mathrm{T}$[HCN(1--0)] & $L_{08}$[HCN(1--0)]  & $f_{08}$ &
$\alpha_{08}$(HCN)  \\
& [$M_\sun$] & [K km s$^{-1}$ pc$^2$] & [K km s$^{-1}$ pc$^2$] & &
[$M_\sun$(K km s$^{-1}$ pc$^2$)$^{-1}$] 
\\
\noalign{\smallskip}
\hline
\noalign{\smallskip}
California & 4,800 & 210 & 16 & 0.08 & 23 \\
Perseus & 5,600 & 77 & 28 & 0.37 & 73 \\
Orion A & 24,000 & 524 & 290 & 0.55 & 46 \\
\noalign{\smallskip}
\hline
\end{tabular}
\tablefoot{
\tablefoottext{a}{$M_{08}$ represents the mass of gas
over the $A_\mathrm{K} = 0.8$ mag 
threshold, $L_\mathrm{T}$ the total line luminosity, $L_{08}$ the line 
luminosity from gas over the $A_\mathrm{K} = 0.8$ mag threshold, 
$f_{08}=L_{08}/L_\mathrm{T}$, and $\alpha_{08}$ the dense gas conversion 
factor (=$M_{08}/L_\mathrm{T}$)}.
}
\end{table*}

The low (and variable) contribution from the high density gas to the
total HCN luminosity of California, Perseus, Orion A, and other clouds
shows that if the HCN luminosity is proportional to the
amount of star-forming gas in a cloud, it is not because HCN traces that gas
directly, but may be because it acts as
a proxy of the star-forming material \citep{don23}.
With this caveat in mind, we determined the amount of dense gas ($M_{08}$)
in each cloud 
by integrating the H$_2$ column density 
over the 0.8~mag threshold in the
maps of \cite{lom14}, \cite{zar16}, and \cite{lad17}, and assuming a
solar value for the metallicity \citep{asp21}. Dividing the derived dense-gas 
masses by the HCN(1--0) luminosities, we estimated the
$\alpha_{08}$(HCN) factors reported in Table~\ref{tbl_alpha_glbl}.
As can be seen,
our $\alpha_{08}$(HCN) estimates span more than a factor of 3, and
range (in units of $M_\sun$(K km s$^{-1}$ pc$^2$)$^{-1}$)
from 23 for California to 73 in Perseus, with Orion A having
an intermediate value of 46.
We note that our estimate for Perseus is very close to the 
76~$M_\sun$(K km s$^{-1}$ pc$^2$)$^{-1}$ derived by \cite{dam23}
when these authors take into account the contribution of the HCN emission that
lies outside their mapping boundary. Our value for Orion A, on the other hand, is 
more than a factor of 2 higher than the value estimated by \cite{kau17},
most likely due to a different estimate of the HCN luminosity, while
no determination of $\alpha_{08}$(HCN) for California 
had so-far been presented.
Taken together, our estimates suggest that $\alpha_{08}$(HCN) varies between
clouds, and that no single HCN conversion factor can be used as a reference
value. A similar diversity of conversion factors has been found by \cite{eva20} and 
\cite{pat22}. In our case, all values are significantly larger than the 
canonical 10~$M_\sun$(K km s$^{-1}$ pc$^2$)$^{-1}$ derived by \cite{gao04}
and commonly used in extragalactic work \citep{use15,gal18,jim19}.

While our sample of three clouds is too small to investigate in general the
origin of the $\alpha_{08}$(HCN) variations, it already offers some clues
on what cloud properties are likely to affect the value of $\alpha_{08}$(HCN).
A first property to consider is the gas temperature, which we have seen in
previous sections significantly influences the intensity of most molecular 
lines.
A dependence of $\alpha_{08}$(HCN) on temperature was already predicted by
\cite{gao04}, who derived a $1/T_{\mathrm{b}}$ scaling from their original
estimate.
While these authors used  
the line brightness temperature as a parameter instead of the
more physical gas kinetic temperature, it is expected that the two will
be related even if the lines are not fully thermalized.
In this regard, it is interesting to note that \cite{gao04} 
assumed an HCN(1--0) brightness temperature of 35~K, which may have been correct
for some of the ultra-luminous IR galaxies of their sample, but is
clearly too large for the galactic clouds of our study.
This can seen from Fig. B.2 in \citealt{taf21}, which shows that the HCN(1--0)
line toward the densest parts of Perseus reaches a brightness temperature that is
a full order of magnitude lower than assumed by \cite{gao04}. As a result,
even a simple application of the $T_{\mathrm{b}}$ scaling law predicts a
conversion factor close to 100~$M_\sun$(K km s$^{-1}$ pc$^2$)$^{-1}$,
which is closer to the value we derive for Perseus.
Since we now know that the \cite{gao04} estimate is too simple an approximation,
it is critical to determine how $\alpha$(HCN) depends on temperature with real data.
While this cannot be done using our limited sample of 
clouds, it can be investigated using the
local version of $\alpha$(HCN), and for
this reason, we defer further discussion of the temperature effects to the 
next subsection.

Another parameter that affects the 
value of $\alpha_{08}$(HCN) and can cause variations 
between clouds is the contribution from the outermost cloud layers. 
As illustrated in Fig.~\ref{fig_hcn_lum}, these layers contribute significantly 
to the total HCN luminosity, but since they do not contribute to the amount 
of dense
gas, their net effect is to decrease the value of $\alpha_{08}$(HCN). 
The most extreme example of this effect is seen in
the California cloud, where about 92\% of the HCN luminosity emerges
from regions below the $A_\mathrm{K}=0.8$~mag dense-gas threshold
(the fraction is 63\% in Perseus and 45\% in Orion A; see 
Table~\ref{tbl_alpha_glbl}).
Not surprisingly,
California presents the lowest $\alpha_{08}$(HCN) value of our
sample (23~$M_\sun$(K km s$^{-1}$ pc$^2$)$^{-1}$),
which is about half that of Orion A and one third of that of
Perseus.
While California may represent an extreme example of a cloud
in terms of its diffuse structure \cite{lad17}, Perseus and
Orion A also present significant differences in terms
of the contribution from their outer layers to the HCN luminosity,
an effect also seen by \cite{eva20} toward clouds in the inner
galaxy. While it may be possible that any cloud-to-cloud
differences average out in extragalactic observations 
that contain multiple clouds inside a telescope bin,
it is likely that a multiplicity of $\alpha_{08}$(HCN)
values is intrinsic to any cloud population.

We finish our analysis of the global $\alpha_{08}$(HCN) factor by noting that
the potentially large contribution from the outer layers 
to the HCN(1--0) luminosity imposes a serious difficulty in estimating 
HCN(1--0) luminosities and therefore $\alpha_{08}$(HCN) factors
both to the mapping and sampling techniques. 
Our observations of Perseus and Orion A show that there is 
residual HCN(1--0) emission 
even in the lowest column density bins of these clouds, whose extinction is
in the range to $A_{\mathrm{V}} \approx$ 1-2~mag. 
Any observation
that does not sample this low-extinction regime runs therefore the risk of
underestimating the total luminosity and therefore overestimating the 
$\alpha_{08}$(HCN) factor. For the California cloud, the
HCN(1--0) line was not detected in the lowest column density bin, and from 
the rms level of the average spectrum we estimate that the
possible contribution from this bin is on the order of 10\% of
the total luminosity, although the figure is clearly uncertain.
A shown in Sect.~\ref{sec_lum}, our luminosity estimate of 
the more extended CO(1--0) emission matches the independent
estimate of \cite{lew22}, so we have some confidence that the sampling 
technique can provide a meaningful estimate of the HCN luminosity
even if dominated by the cloud outermost layers.

\subsubsection{The local $\alpha_\mathrm{X}$(HCN) factor}

\begin{figure*}
        \centering
        \includegraphics[width=0.8\hsize]{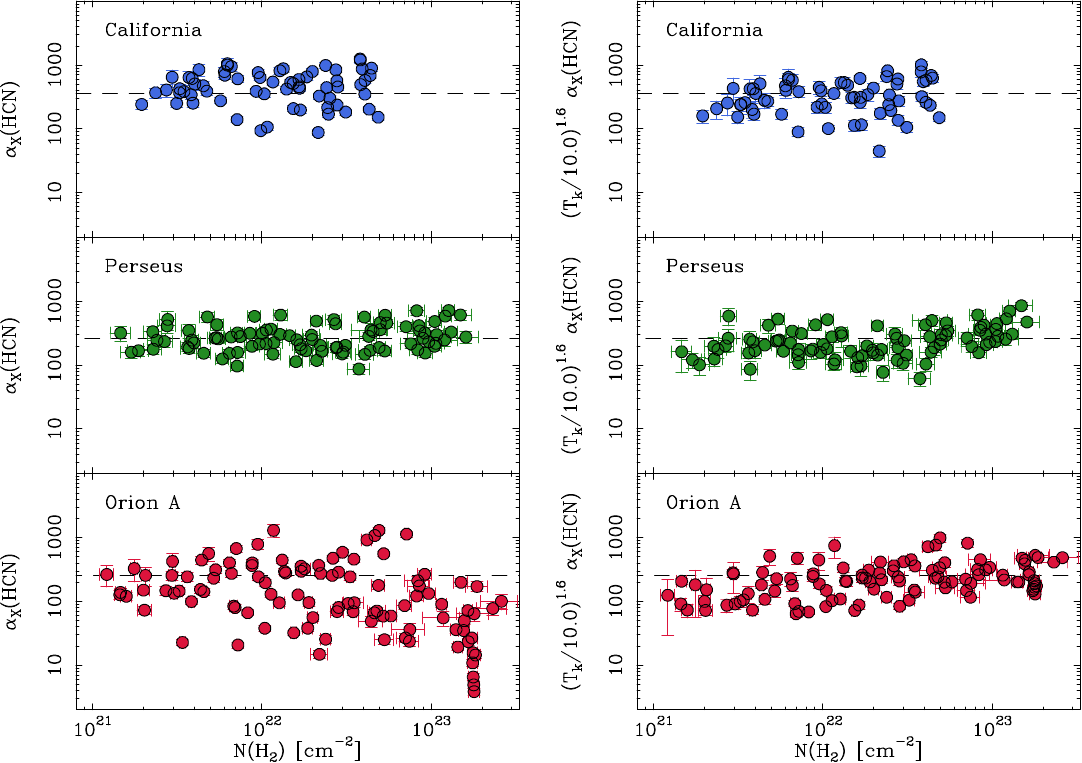}
        \caption{Local $\alpha_{\mathrm{X}}$(HCN) factor as a function of H$_2$
        column density for the California, Perseus, and Orion A clouds.
        {\em Left:} $\alpha_{\mathrm{X}}$(HCN) factor calculated using
         Eq.~\ref{eq_alpha_x}.
        {\em Right:} $\alpha_{\mathrm{X}}$(HCN) factor multiplied by the
        temperature factor ($T_{\mathrm{k}}$/10~K)$^{1.6}$ to compensate
        for the dependence determined in Eq.~\ref{eq_alpha_temp}.
        Note the reduced dispersion of the Orion A data.        
        In all panels, the units of $\alpha_{\mathrm{X}}$(HCN) are in 
        M$_\odot$~(K km s$^{-1}$ pc$^2$)$^{-1}$, and
        the dashed lines represent the mean 
        value of the temperature-corrected factor for  each cloud.}
        \label{fig_alpha}
\end{figure*}

We now investigate the information about the HCN(1--0) emission as a dense-gas tracer that can be derived from the local definition of the conversion factor. 
This definition follows common practice in the 
analysis of the CO emission as a cloud tracer, where the term conversion factor 
is referred indistinctly to both the ratio between H$_2$
column density and integrated intensity (represented by $X$), and the 
ratio between total cloud mass and line luminosity (represented by $\alpha$; 
see \citealt{bol13} for a review).
Following this convention, we define
\begin{equation}
\alpha_{\mathrm{X}}(\mathrm{HCN}) \; [\mathrm{M}_\odot~(\mathrm{K~km~s}^{-1} 
\mathrm{~pc}^2)^{-1}]  
= 2.25 \times 10^{-20} \; 
\frac{N{\mathrm{(H}}_2\mathrm{)\; [cm}^{-2}]} 
{I_{\mathrm{HCN}} \; \mathrm{[K~km~s}^{-1}]},
\label{eq_alpha_x}
\end{equation}
where we have assumed a solar abundance of the elements \citep{asp21}
to convert the H$_2$ column density into a mass density,
and used the subindex X following the convention proposed by \cite{dam23}
(see, e.g., Eq.~7 in \cite{eva22} for an equivalent definition using CO).
As expected from Eq. 4,
the global and local conversion factors are closely related: the
global factor corresponds to the intensity-weighted cloud average of 
the local factor
after substituting the total H$_2$ column density
by the column density of dense gas.

The local $\alpha_\mathrm{X}$(HCN) 
factor has been used by \cite{shi17}
to investigate the relation between the HCN emission and the gas mass
in the Aquila, Ophiuchus, and Orion B clouds (see their Fig.~6).
Our sampling observations provide a natural data set to carry out a similar
investigation in California, Perseus, and Orion A  since the 
ratio between the H$_2$ column density and the HCN(1--0) intensity
can be directly determined from the sampling data.
The left panels of Fig.~\ref{fig_alpha} show the distribution of
$\alpha_{\mathrm{X}}$(HCN) as a function of $N$(H$_2$) for
California, Perseus, and Orion A when no
temperature correction has been applied to the HCN emission.
As can be seen, the $\alpha_{\mathrm{X}}$(HCN) parameter
remains approximately constant in California and Perseus,
as expected from the close-to-linear
dependence of the HCN(1--0) intensity on $N$(H$_2$) found in 
Sect.~\ref{sect_trad_dense}. In addition, it
presents relatively low levels of dispersion of 0.27 and 
0.21 dex, respectively.
The $\alpha_{\mathrm{X}}$(HCN) parameter in Orion A, on the other hand,
remains approximately constant for $N$(H$_2$)
lower than $10^{23}$cm$^{-2}$ but drops significantly
at higher column densities and presents a 
higher level of scatter of about 0.53 dex.
Overall, our $\alpha_{\mathrm{X}}$(HCN) distributions look similar
to those found by \cite{shi17} in Aquila, Ophiuchus, and Orion B,
which show close-to-constant dependence on extinction.
The Perseus distribution, in addition, presents a similar average to
that determined by \cite{dam23} for this cloud (=215),
as would have been expected from their similar estimate of the $X$ factor.

\begin{figure}
        \centering
        \includegraphics[width=\hsize]{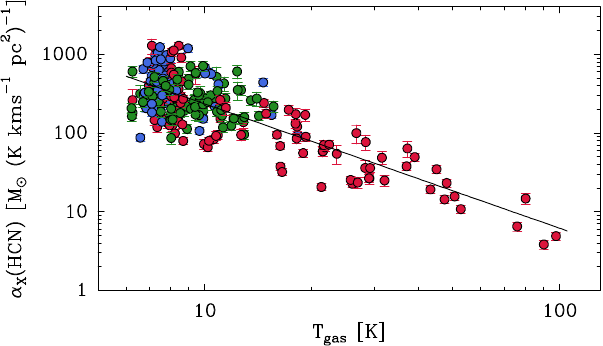}
        \caption{$\alpha_{\mathrm{X}}$(HCN) factor as a function of gas
        temperature in California, Perseus, and Orion A.
        The solid line represents the fit described in the text.
        The data are color-coded as in previous figures.}
        \label{fig_alpha_temp}
\end{figure}

To investigate the origin of the higher dispersion of $\alpha_{\mathrm{X}}$(HCN)
in Orion A, we looked at the dependence of the HCN emission on temperature.
Section~\ref{sect_trad_dense} and 
previous studies of the HCN(1--0) emission in Orion A
have found a systematic dependence of the emission on temperature
(\citealt{gol81,sch92,gra14}), and \cite{hac20}
estimated that the ratio between HCN(1--0) intensity 
and visual extinction depends quadratically on the gas kinetic temperature 
up to 40~K (these authors excluded from their analysis 
the hottest vicinity of the ONC).
Figure~\ref{fig_alpha_temp} presents our estimate of the dependence of the
$\alpha_{\mathrm{X}}$(HCN) factor on
gas temperature for the combined data from California, Perseus, and Orion A.
As can be seen, at low temperatures ($\sim 10$~K) the data from the
three clouds shows a significant overlap, while at higher temperatures
(> 15~K), which 
are only represented by the Orion A data, $\alpha_{\mathrm{X}}$(HCN)
systematically decreases with temperature
by almost two orders of magnitude. From a linear fit to the log-log plot
we determine that $\alpha_{\mathrm{X}}$(HCN) 
depends on the gas kinetic temperature as
\begin{equation}
\alpha_{\mathrm{X}}(\mathrm{HCN}) \;  [\mathrm{M}_\odot~(\mathrm{K~km~s}^{-1} 
\mathrm{~pc}^2)^{-1}]   = 235\pm 41\; 
\left(\frac{T_\mathrm{gas}} {10~\mathrm{K}}\right)^{-1.6\pm0.1}.
 \label{eq_alpha_temp}
\end{equation}This dependence of $\alpha_{\mathrm{X}}$ on the gas kinetic temperature is 
steeper than the $T_{\mathrm b}^{-1}$ predicted by \cite{gao04} (assuming a close relation between the brightness and kinetic temperatures).
This is likely the result from a combination of our more realistic dependence
of the line intensity on the gas temperature (Appendix~\ref{app_temp}) and the
high sensitivity of the HCN abundance with gas temperature found in
Sect.~\ref{sec_rare}.

The systematic correlation of $\alpha_{\mathrm{X}}$(HCN) with gas temperature
suggests that the peculiar behavior of the Orion A data in Fig.~\ref{fig_alpha}
results from the temperature variations in the cloud. 
To test this idea, we multiplied the $\alpha_{\mathrm{X}}$(HCN) factor
by ($T_\mathrm{k}$/10~K)$^{1.6}$, which is the inverse of the power law derived 
from the fit, and present the result in the right panels of Fig.~\ref{fig_alpha}.
As can be seen, the 
temperature-corrected conversion factor in Orion A shows an
approximately constant dependence on $N$(H$_2$) and has a similar dispersion to
that measured in California and Perseus (0.2-0.3 dex). A slight drop of 
$\alpha_{\mathrm{X}}$(HCN) al low $N$(H$_2$)
in the temperature-corrected  values of 
Perseus and Orion~A likely results from small errors in the 
temperature at low column densities 
when using the dust temperature as a reference
(Sect.~\ref{sect_temp}). 

The similar distribution of the corrected  $\alpha_{\mathrm{X}}$(HCN) 
factors in the three clouds
suggests that gas temperature differences were responsible 
for the observed differences in the uncorrected factors.
This interpretation differs from that of  
\cite{shi17}, who also found differences in the conversion factor
between their clouds, but associated them with variations in the local far-UV
(FUV) radiation field, which they estimated to range 
from $G_0=1$ to more than 4000.
Interpreting the $\alpha_{\mathrm{X}}$(HCN) differences as a result of
the FUV radiation, however, presents several problems. 
First of all, it is unlikely that the HCN-emitting gas is directly
exposed to the high levels of FUV radiation measured toward the 
exterior of the clouds
since UV radiation quickly photodissociates the HCN molecules \citep{agu17}.
In addition, a dependence of $\alpha_{\mathrm{X}}$(HCN) 
on the external FUV radiation field seems to contradict the observed 
constant behavior of this factor as a function of column density 
since the FUV radiation is expected to be
strongly attenuated by the cloud internal extinction.
It is therefore more likely that the cloud-to-cloud variations
found by \cite{shi17} also arise from differences in the cloud gas temperature.
This is further supported by the fact that \cite{shi17} 
used the dust temperature to infer the $G_0$ factor, so there is a
possible ambiguity interpreting the effect of the two parameters.

If the gas 
temperature has the strong effect on the local $\alpha_{\mathrm{X}}$(HCN)
factor suggested by Eq.~\ref{eq_alpha_temp}, a similar 
dependence on temperature 
is expected to affect the global
factor, which we have seen represents an
intensity-weighted average of $\alpha_{\mathrm{X}}$(HCN) over a whole cloud. 
This temperature dependence may be difficult to observe in nearby
galactic clouds because the contribution from relatively warm regions
will likely be overwhelmed by the contribution from
the more extended colder gas that we have seen dominates the HCN emission.
In extragalactic observations, on the other hand, it may be possible to
encounter more extreme conditions where the warm gas dominates 
the global HCN emission. Indeed, observations
of luminous and ultra-luminous infrared galaxies suggest that these systems have significantly
lower conversion factors than normal galaxies \citep{gar12},
as expected from their elevated gas temperatures.
This result should serve as a warning that 
for neither the local nor the global versions 
of the conversion factor, one size fits all, and that 
care must be exercised when applying the same conversion
factor to inhomogeneous samples of clouds or galaxies.

\subsection{The HCN/CO ratio and its correlation 
with the H$_2$ column density}

\begin{figure}
        \centering
        \includegraphics[width=\hsize]{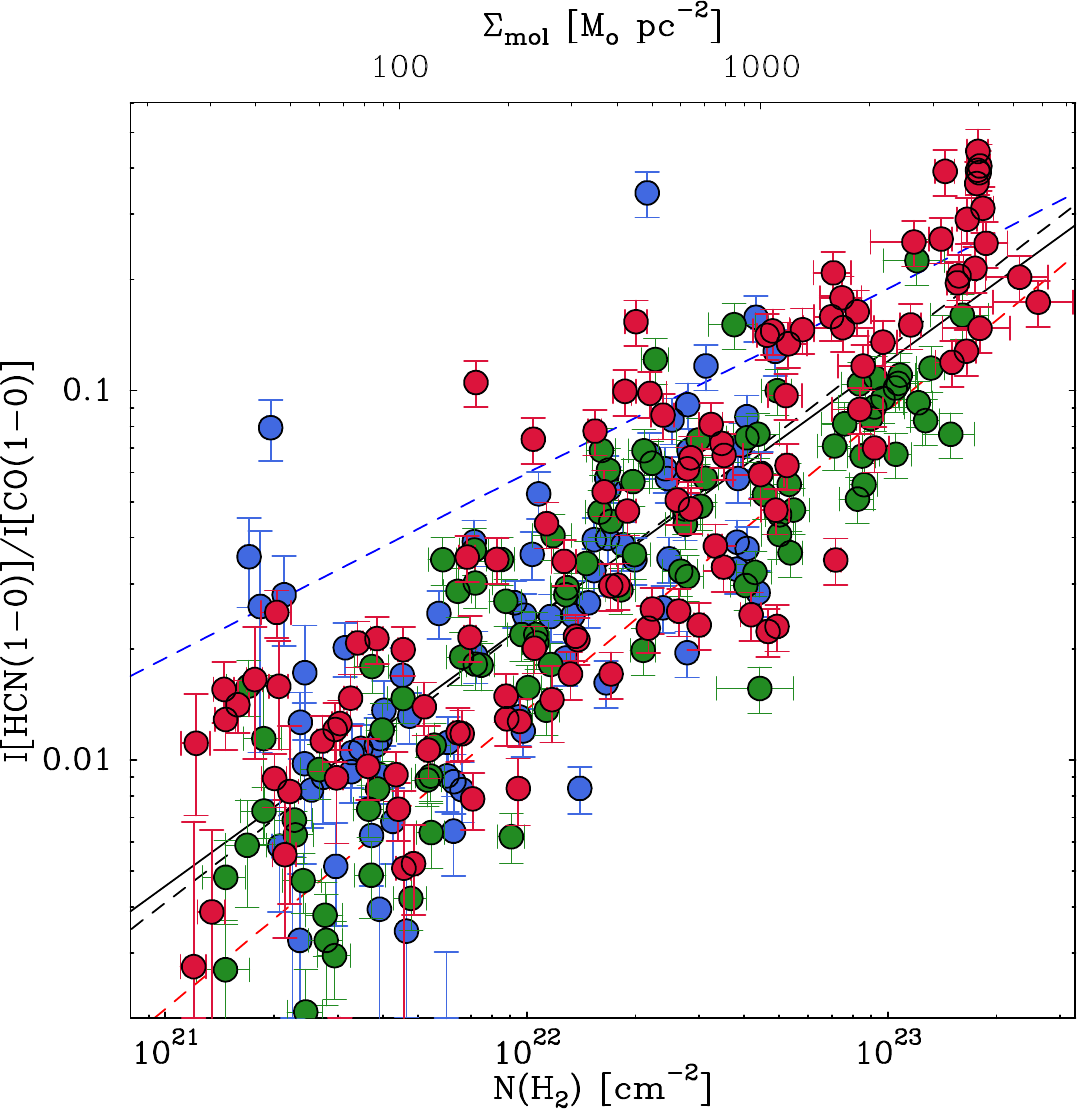}
        \caption{HCN(1--0)/CO(1--0) intensity ratio as a function of H$_2$ 
        column density for California, Perseus, and Orion A, color-coded 
        as in Fig.~\ref{fig_co1}. The solid 
        and dashed black lines represent the two fits discussed in the text. The 
        dashed red and blue lines represent, respectively, the fits derived 
        by \cite{gal18} and \cite{jim19} from extragalactic data.}
        \label{fig_hcn_co}
\end{figure}

Another parameter commonly used to interpret extragalactic observations 
is the HCN/CO intensity ratio.
Assuming that the HCN intensity is proportional to the 
dense gas column density, and that the CO intensity is proportional
to the total gas column density, the HCN/CO intensity ratio is
expected to measure the fraction of dense gas 
\citep{gao04,use15,ler17}.
Recent work by \cite{gal18} and \cite{jim19}
has found a significant correlation between the HCN/CO ratio and 
the molecular column density in (normal) galaxies averaged
over 1-2 kpc spatial scales.
\cite{gal18} have interpreted this correlation as indicating that both the
HCN/CO ratio and the gas column density are sensitive tracers 
of the density distribution in molecular clouds.

Our line survey provides estimates of the HCN and CO intensity 
together with the $N$(H$_2$) column density, so we can use the data 
to investigate the correlation between 
the HCN/CO ratio and $N$(H$_2$) in galactic clouds.
Figure~\ref{fig_hcn_co} presents the HCN(1--0)/CO(1--0) intensity ratio
(HCN/CO hereafter)
for the three clouds of our survey 
as a function of both the H$_2$ column density
(lower x-axis) and the molecular surface density commonly used in  
extragalactic work 
(upper x-axis).\footnote{$\Sigma_{\mathrm{mol}}\; [\mathrm{M}_\sun
\; \mathrm{pc}^{-2}] =  2.25\times 10^{-20}\; N(\mathrm{H}_2)\; 
[\mathrm{cm}^{-2}]$
assuming a standard solar abundance \citep{asp21}.}
No temperature correction has been applied to either the HCN or CO data,
so the results can be directly compared with extragalactic observations. 
Applying a
temperature correction to the two lines, however, only has a minor
effect on the intensity ratio since the two corrections almost cancel out
and the resulting scatter plot is practically indistinguishable from that of 
Fig.~\ref{fig_hcn_co}.
As can be seen in the figure, the HCN/CO ratio correlates strongly
with the H$_2$ column density over the
more than two orders of magnitude covered by this parameter.
In addition, the correlation seems to be the same in the three clouds,
an impression confirmed by FF tests of the three possible
cloud pairs, which return p-values between 0.28 and 0.76.
Combining the data from the three clouds, we estimate a  Pearson's coefficient
of 0.84 (in log-log scale), and
using a least squares fit we derive a relation of the form
\begin{equation}
\log_{10} \left( \frac{I_\mathrm{HCN}}{I_\mathrm{CO}} \right) = (-3.3 \pm 0.8)
+ (0.71 \pm 0.03) \; \log_{10} \left( \frac{\Sigma_{\mathrm{mol}}}{{\mathrm{M}_\sun \; \mathrm{pc}^{-2}}}\right),
\label{eq_hcn_co}
\end{equation}
where the intensities refer to the $J$=1--0 transition of both HCN and CO.
This fit is represented in the figure by a black solid line.

The correlation between HCN/CO and $N$(H$_2$) found in our three clouds
is remarkably similar to that seen at kiloparsec scales in external galaxies
by \cite{gal18} and \cite{jim19}.
These extragalactic observations cover similar ranges of $N$(H$_2$) as our cloud data
($10^2$-$10^3$~M$_\odot$~pc$^{-2}$ for
\citealt{gal18} and 10-300~M$_\odot$~pc$^{-2}$ for \citealt{jim19},
although these values include
the contribution of the filling factor of the clouds), and are
illustrated in Fig.~\ref{fig_hcn_co} using red and blue dotted lines.
As the plot shows, 
our galactic fit, with a slope of $0.71\pm 0.03$, is 
intermediate between the fits obtained by \cite{gal18}
(slope $0.81\pm 0.09$) and \cite{jim19} (slope $0.5\pm 0.1$),
who used different assumptions
to estimate the molecular surface density.

Further work is needed to better connect the galactic and extragalactic 
results,
both in terms of the disparate spatial scales that they sample 
(subparsec and 1-2~kpc, respectively) and the different methods
used to derive the
H$_2$ column density, which in the extragalactic case relies on
indirect uses of the CO emission \citep{gal18,jim19}.
Assuming that the different estimates are truly comparable,  
the most natural interpretation of the similar behavior of the HCN/CO ratio
is that the extragalactic correlation reflects 
the internal properties of the individual unresolved clouds. 
To explore how these properties could give rise to the 
HCN/CO versus $N$(H$_2$) correlation seen in our data set, we needed to
use the results from the 
radiative transfer model presented in Paper I to reproduce the Perseus data. 
The similar behavior of the HCN/CO ratio in the three clouds of our sample
suggests that the
excitation mechanism responsible for the Perseus correlation is likely also
responsible for the California and Orion A correlations.

According to Paper I, the intensity of multiple
transitions, including CO(1--0) and HCN(1--0), can be reproduced with a model that 
assumes that the gas physical and chemical properties depend on $N$(H$_2$)
(see Table 3 in Paper I).
Of particular interest for the HCN/CO correlation is the relation between
the volume density and the column density, which was found to have the form 
$n(\mathrm{H}_2) = 2 \times 10^4 \mathrm{cm}^{-3}\;  
(N$(H$_2)/10^{22} \mathrm{cm}^{-2})^{0.75}$. 
Using this relation, 
the radiative transfer model of Paper I showed that 
both CO(1--0) and HCN(1--0) must be  optically thick
over most of the cloud, and that while
CO(1--0) is thermalized at all column densities, HCN(1--0) remains 
sub-thermal with an excitation temperature strongly dependent on $N$(H$_2$)
(see Fig.~13 in Paper I).
This excitation 
behavior makes the intensity of HCN(1--0) rapidly increase 
with $N$(H$_2$) (through its density dependence),
 while the intensity of CO(1--0) stays approximately constant.
As a result, the HCN/CO ratio systematically increases with $N$(H$_2$), in 
agreement with the observed behavior.

To next interpret the HCN/CO ratio as an indicator of the gas volume density,
we combined the relation between volume and column densities derived in Paper I 
with our fit of the HCN/CO data.
As seen in Eq.~\ref{eq_hcn_co}, the HCN/CO ratio depends on  $N$(H$_2$)
with a slope of $0.71\pm 0.03$, which differs by only 1.3 $\sigma$ from 
the 0.75 value determined for the density relation with $N$(H$_2$).
Taking this similarity of values as an indication of an approximate equality, we re-fitted the HCN/CO-$N$(H$_2$) correlation using a 
fixed value of 0.75. The result is represented in Fig.~\ref{fig_hcn_co}
with a dashed line, and is practically indistinguishable from the original best 
fit inside the range of values covered by the observations. 
Using this new fit (which has an intercept of -3.4), we 
derive a relation between gas density and the HCN/CO ratio of the form
\begin{equation}
n(\mathrm{H}_2) = 8.7\times 10^5\; \mathrm{cm}^{-3}\; 
\frac{I_\mathrm{HCN}} {I_\mathrm{CO}}.
\label{eq_dens}
\end{equation}
This volume density should be interpreted as a mean value along the
line of sight where the HCN/CO ratio has been measured, and since its derivation
uses the radiative transfer model of Paper I, the mean
has been weighted by the emission of the CS and HCN.
No meaningful error bar could be estimated for this density due to the 
difficulty in quantifying the model assumptions, but given
the quality of the
model fits, it is likely that the uncertainty lies within a 
factor of 2.  

It is probably premature to extrapolate
our derived relation between volume density and HCN/CO ratio 
to extragalactic data, although the strong 
similarity between the galactic and extragalactic correlations of
HCN/CO with $N$(H$_2$) suggests that this is likely to be the case.
If so, the HCN/CO ratio should be thought not so much as an indicator of
the dense gas fraction but as an estimator of the gas 
volume density averaged over the line of sight and the observing beam.
Such an estimator presents
the advantage over single-line tracers that is less sensitive to gas
temperature variations given the approximate cancellation between  
the dependences of HCN and CO.
Further characterization of the HCN and CO emission from galactic clouds 
is needed to study the general properties of the line ratio and to
better calibrate its dependence on $N$(H$_2$). For this investigation, 
the stratified random sampling technique presented here
appears to be a suitable tool.

\section{Conclusions}

We sampled the 3 mm wavelength emission
of the California and Orion A clouds
using the IRAM 30 m radio telescope. We selected
a set of target positions using
the stratified random sampling technique previously used in \cite{taf21}
to study the emission from the Perseus cloud. This technique divides
the cloud into multiple bins of H$_2$ column
density and randomly selects a number of cloud positions in each bin
to carry out the molecular-line observations.
We combined the new results from California and Orion A 
with the 
Perseus cloud data to investigate the main gas parameters that control the
line emission of the CO isotopologs and the main dense-gas tracers,
and to compare the emission of these species in three clouds whose star-formation rates
span more than one order of magnitude.
The main results from our study are the following:

1. In the three target clouds, the
intensity of the studied molecular lines correlates strongly with the 
value of the H$_2$ column density even if the positions 
are separated by distances of tens of parsecs.
This strong correlation with $N$(H$_2$) shows that this parameter is 
the main predictor of the line
intensity and supports its use in the stratified random sampling technique.

2. The observations of Orion A, which presents gas temperature variations
across its components, show that the
intensity of most molecular lines also depends on the
gas temperature. We used a cloud radiative transfer model 
to determine the expected change in the intensity of all target lines as a
function of gas temperature, and used the model results to simulate
the emission expected 
if our target clouds were isothermal. The temperature-corrected  
intensities present a lower level of dispersion and
a better agreement between the three target clouds
than the uncorrected intensities. 

3. We find that the temperature-corrected
intensity of the CO lines has a flatter-than-linear
dependence on $N$(H$_2$), while the intensity of traditional dense-gas tracers such as HCN(1--0), CS(2--1), HCO$^+$(1--0), and HNC(1--0) 
scales almost linearly with $N$(H$_2$)
over the two orders of magnitude covered by the observations
($\approx 10^{21}$-$10^{23}$~cm$^{-2}$).

4. In contrast with the traditional dense-gas tracers, the intensity of 
N$_2$H$^+$(1--0) 
does not correlate linearly with $N$(H$_2$) over the full column 
density range. It correlates almost linearly at high column densities, 
but it drops by more than one order of magnitude between 
$2 \times 10^{22}$~cm$^{-2}$ and $10^{22}$~cm$^{-2}$, and remains undetected
at lower column densities. This behavior, which is similar
in the three clouds, makes N$_2$H$^+$ the only selective tracer 
of the cloud cold dense component.

5. In addition to affecting the molecular excitation, 
the gas kinetic temperature 
changes the abundance of some species. Using the intensity distribution of  
rare isotopologs, we find that
the abundance of HCN and CS is systematically enhanced
with increasing gas temperature,
while the abundance of HCO$^+$ and HNC remains approximately constant 
between 10 and 100~K. 
In contrast with the classical dense-gas tracers, N$_2$H$^+$ decreases 
in abundance with temperature, most likely due to the release of
CO from the grains as the temperature increases.

6. The stratified random sampling data can also
be used to estimate cloud-integrated luminosities of the different molecular lines. 
We compared our estimated luminosities with literature values (mostly from
CO isotopologs) and find an agreement typically at the 25\% level, which is
remarkable because the comparison involves very different telescopes 
and calibration schemes.
We used our sampling 
data to estimate luminosities of the main survey lines 
in California, Perseus, and Orion A.
 
7. The systematic emission patterns found in our survey suggest that
the target molecular clouds share a common chemical structure.
This structure is characterized by
abundance variations as a function of column density and can be
approximately understood as consisting of three main chemical regimes.
Between the photodissociation boundary ($\sim 10^{21}$~cm$^{-2}$)
and $\sim 10^{22}$~cm$^{-2}$, most species maintain a 
close-to-constant abundance
that is likely determined by gas-phase reactions. In this regime, N$_2$H$^+$
remains undetected due to the high CO abundance in the gas phase.
From  $\sim 10^{22}$~cm$^{-2}$ to  
$\sim 10^{23}$~cm$^{-2}$, the abundance of most species decreases due to freeze-out onto grains, while N$_2$H$^+$ is enhanced as a result of a decrease in 
the gas-phase CO abundance.
At elevated temperatures and 
column densities higher than  $\sim 10^{23}$~cm$^{-2}$, which in our sample 
are only
reached toward Orion A, high-mass star-formation feedback
disturbs the gas chemical composition, enhancing species such as HCN and CS
and destroying N$_2$H$^+$.

8. We used our survey data to study the relation between the HCN(1--0) emission
and the cloud gas mass. We explored the clues provided by 
two possible definitions of the HCN conversion factor previously used in the literature.
The ``global'' definition compares the cloud-integrated line luminosity with the
amount of ``dense'' gas and shows variations of more than a factor of 3
between California, Perseus, and Orion A.
These variations mostly arise from the different contribution to the HCN
emission of the external layers of the cloud, which tend to dominate the 
luminosity due to their large surface area.
A ``local'' definition of the $\alpha$(HCN) 
factor compares the HCN(1--0) intensity with the 
total H$_2$ column density and can be measured at each cloud position.
This factor displays a strong dependence on the gas kinetic temperature,
which seems to result from a combination of excitation and abundance effects.
A dependence of the $\alpha$(HCN) factor on the gas temperature
may help explain the diversity of values seen by galactic and extragalactic observers.

9. We also used our survey data to study the correlation between 
the HCN(1--0)/CO(1--0) intensity ratio and the  gas column 
density, which has recently been studied using extragalactic observations.
Our data show a similar relation in terms of both the
range of parameters and slope. 
Using the results of a cloud radiative transfer model, we show
that the HCN(1--0)/CO(1--0) ratio can be used to estimate the 
mean gas volume density, and 
that the correlation with $N$(H$_2$) observed in our clouds results from the 
gradual increase in the HCN(1--0) sub-thermal excitation with H$_2$ column density.

The above results illustrate the great
potential of the stratified sampling technique to
characterize the molecular emission from star-forming clouds.
Given its
relatively low cost in terms of telescope observing time,
it should be possible to expand its application to
a larger number of clouds and obtain a
more complete view of their intrinsic diversity
that can further our understanding of star formation
and can serve as a template to analyze
extragalactic observations.

\begin{acknowledgements}        
We thank our referee, Neal Evans, for a thorough and critical review of the
manuscript that helped us improve the analysis and the presentation, and for
information on the results of \cite{yun21}.
We thank Toshikazu Onishi for valuable information on the calibration scale 
of the Osaka 1.85m telescope.
MT and AU acknowledge partial support from project PID2019-108765GB-I00 
funded by MCIN/AEI/10.13039/501100011033.
AU acknowledges support from the Spanish grant PGC2018-094671-B-I00, 
funded by MCIN/AEI/10.13039/501100011033 and by ``ERDF A way of making Europe''.
This project has received funding from the European
Research Council (ERC) under the European Union’s
Horizon 2020 research and innovation programme (Grant
agreement No. 851435)
This work is based on IRAM 30 m-telescope observations
carried out under project numbers 034-17, 104-17, 033-18, 008-19, and 116-20.
IRAM is supported
by INSU/CNRS (France), MPG (Germany), and IGN (Spain). This research has
made use of NASA's Astrophysics Data System Bibliographic Services and the
SIMBAD database, operated at CDS, Strasbourg, France.   
\end{acknowledgements}

\bibliographystyle{aa}
\bibliography{46136.bib}

\appendix

\section{Maps of sample points for California and Orion A}
\label{app_plots}

\begin{figure*}
        \centering
        \includegraphics[width=0.95\hsize]{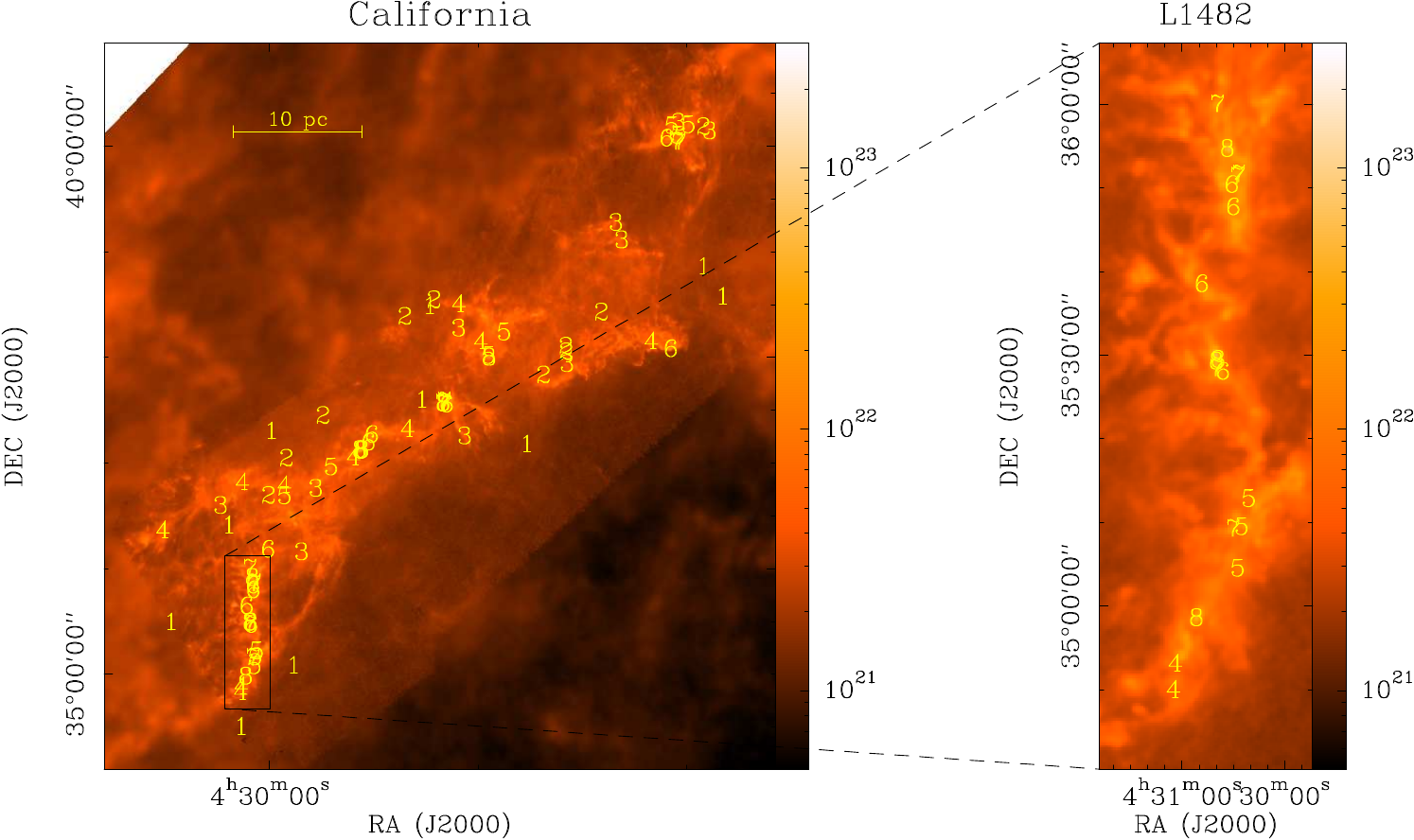}
        \caption{Sampling of the California cloud. The color images represent 
        the distribution of column
        density as derived by \cite{lad17} from {\em Herschel} Space Observatory
        data, and the numbers indicate the positions chosen using stratified
        random sampling. The right panel is an expanded view of the
        L1482 region, which contains some of the highest column density 
        positions.
        The numbers refer to the column density bin of the
        position (1 is lowest and 8 is highest), the 10~pc scale bar assumes a 
        distance of 470~pc \citep{zuc19}, and the wedge scale is in units of 
        cm$^{-2}$.}
        \label{map_calif}
\end{figure*}

\begin{figure*}
        \centering
        \includegraphics[width=0.95\hsize]{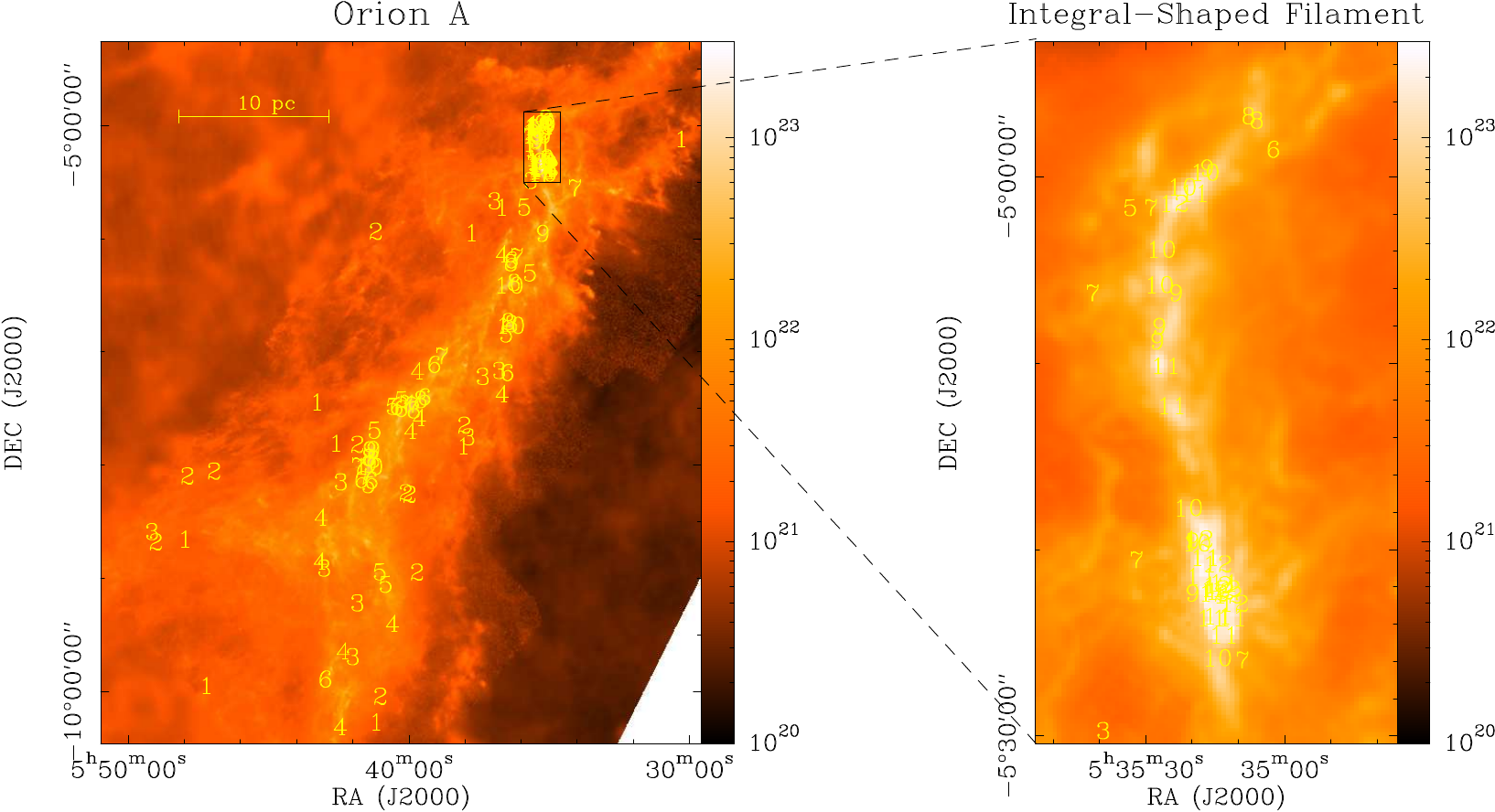}
        \caption{Sampling of the Orion A cloud. The color images represent 
        the distribution of column
        density as derived by \cite{lom14} from {\em Herschel} Space Observatory
        data, and the numbers indicate the positions chosen using stratified
        random sampling. The right panel is an expanded view of the
        ISF, which contains some of the highest column 
        density positions.
        The numbers refer to the column density bin of the
        position (1 is lowest and 12 is highest), the 10~pc scale bar assumes a 
        distance of 432~pc \citep{zuc19}, and the wedge scale is in units of 
        cm$^{-2}$.}
        \label{map_oria}
\end{figure*}

Figures \ref{map_calif} and \ref{map_oria} represent in color scale 
the column density maps of
California and Orion A derived by \cite{lad17} and \cite{lom14}.
The numbers mark the positions selected using
the stratified random sampling technique, and numerical 
values indicate the column density bin to which the position belongs.

\clearpage

\section{Sample positions and 3 mm line intensities}
\label{app_positions}

Tables \ref{tbl_master_calif} and \ref{tbl_master_oria} present the coordinates, H$_2$ column density estimates, 
and 3 mm line
intensities for all the target positions in the California and
Orion clouds. Full versions of
the tables are available at the CDS.

\begin{table*}[h!]
\caption[]{Sample positions and line intensities for California.
\label{tbl_master_calif}}
\centering
\begin{tabular}{lccccccccccc}
\hline\hline
\noalign{\smallskip}
 Position\tablefootmark{a} & RA(J2000) & Dec(J2000) & $N$(H$_2$) & $^{12}
 $CO(1--0) & $^{13}$CO(1--0) & C$^{18}$O(1--0) & HCN(1--0) &  CS(2--1) \\
& ($^\mathrm{h}$~~$^\mathrm{m}$~~$^\mathrm{s}$) & (\degr~~\arcmin~~\arcsec) & 
(cm$^{-2}$) & (K km s$^{-1}$) & (K km s$^{-1}$) & (K km s$^{-1}$) &  (K km 
s$^{-1}$) & (K km s$^{-1}$) \\
\hline
\noalign{\smallskip}
CAL-08\_10 & 04~19~28.8  &  +37~59~04 &  $3.81 (0.04) \times 10^{22}$ &
 21 (2)  & 4.6 (0.5) &   2.1 (0.2) &  0.69 (0.07) &  0.65 (0.07) \\
CAL-08\_09 & 04~25~36.1  &  +37~06~05 & $4.46 (0.05) \times 10^{22}$ &  
19 (2)  & 6.6  (0.7) &  2.3 (0.2) &  1.1 (0.1) & 0.93 (0.09) \\
CAL-08\_08 &  04~25~40.2 &  +37~07~01 &  $3.83 (0.03) \times 10^{22}$ &       
18 (2) &  5.6  (0.6) & 2.4 (0.2) & 0.72 (0.07) & 0.69  (0.07) \\
CAL-08\_07 & 04~30~48.5 & +34~58~03 &  $4.31 (0.14) \times 10^{22}$ &
30 (3)  & 9.9  (1.0) &  1.3 (0.1) &  4.8 (0.5) &  2.8 (0.3) \\
CAL-08\_06  & 04~30~39.4 & +35~29~02  &  $4.08 (0.05) \times 10^{22}$ &      
43 (4) &  17 (2) &  4.3 (0.4) &  1.6 (0.2) & 1.4 (0.1) \\
CAL-08\_05 &  04~25~32.0 & +37~07~01 & $3.92 (0.02)  \times 10^{22}$ &
14 (1) & 4.3 (0.4) & 1.5 (0.2) &  1.0 (0.1) &  0.50 (0.05) \\
CAL-08\_04  & 04~30~36.8 & +35~54~03 &  $4.88 (0.12) \times 10^{22}$ &       
56  (6) &  17 (2) &  3.6  (0.4) &  7.2 (0.7) & 6.0 (0.6) \\
CAL-08\_03 & 04~21~41.0 & +37~33~05 & $4.07 (0.07) \times 10^{22}$ &        
31 (3) &  8.5  (0.9)  & 2.5 (0.3) &  2.6 (0.3) &  2.6 (0.3) \\
CAL-08\_02 & 04~30~40.4 &  +35~29~01 & $4.37 (0.04)  \times 10^{22}$ &       
50 (5) &  19 (2) &  4.7 (0.5) &  1.4  (0.1) & 1.4  (0.1) \\
CAL-08\_01 & 04~21~42.0 & +37~33~04 & $3.84 (0.07) \times 10^{22}$ &       
30 (3) &  7.1 (0.7) & 2.3 (0.2) & 1.8 (0.2) &  2.3 (0.2) \\  
\hline
\noalign{\smallskip}
\end{tabular}
\tablefoot{A full version of this table is available at the CDS.
\tablefoottext{a}{The first number in the position name indicates the column density bin and the second one indicates the order in our sampling sequence.
Numbers in parentheses indicate 1$\sigma$ errors.}
        }      
\end{table*}

\begin{table*}[h]
\caption[]{Sample positions and line intensities for Orion A.
\label{tbl_master_oria}}
\centering
\begin{tabular}{lccccccccccc}
\hline\hline
\noalign{\smallskip}
Position\tablefootmark{a} & RA(J2000) & Dec(J2000) & $N$(H$_2$) & $^{12}
$CO(1--0) & $^{13}$CO(1--0) & C$^{18}$O(1--0) & HCN(1--0) &  CS(2--1) \\                
& ($^\mathrm{h}$~~$^\mathrm{m}$~~$^\mathrm{s}$) & (\degr~~\arcmin~~\arcsec) & 
(cm$^{-2}$) & (K km s$^{-1}$) & (K km s$^{-1}$) & (K km s$^{-1}$) &  (K km 
s$^{-1}$) & (K km s$^{-1}$) \\
\hline
\noalign{\smallskip}
ORIA-12\_10 & 05~35~14.9 & -06~37~50 & $1.8 (0.1) \times 10^{23}$  & 1380 (140) & 124 (12) & 14.8 (2)  & 611 (62)   &  110 (11) \\
ORIA-12\_09 & 05~35~12.7 & -06~37~50 & $1.8 (0.1) \times 10^{23}$  &  916 (92)  & 73 (7)   & 7.4E (0.7)& 285 (29)   &  47 (5) \\
ORIA-12\_08 & 05~35~14.5 & -06~39~10 & $2.3 (0.4) \times 10^{23}$  & 333 (33)   & 56 (6)   & 5.2 (0.5) & 68 (7)     &  30 (3) \\
ORIA-12\_07 & 05~35~14.5 & -06~38~03 & $1.8 (0.1) \times 10^{23}$  & 1000 (100) & 84 (8)   & 8.5 (8)   & 365 (37)   &  59 (6) \\
ORIA-12\_06 & 05~35~10.9 & -06~37~03 & $1.8 (0.4) \times 10^{23}$  & 428  (43)  & 52 (5)   & 4.0 (0.4) & 63 (6)     & 25 (2) \\
ORIA-12\_05 & 05~35~14.0 & -06~37~43 & $1.8 (0.1) \times 10^{23}$  & 258 (26)   & 172 (17) & 14 (1)    & 1050 (110) & 220 (22) \\
ORIA-12\_04 & 05~35~15.4 & -06~37~36 & $1.8 (0.1) \times 10^{23}$  & 2100 (220) & 136E (14) & 15 (2)   & 816 (82)   & 133 (13) \\
ORIA-12\_03 & 05~35~18.5 & -06~40~31 & $2.6 (0.6) \times 10^{23}$  & 337 (34)   & 57 (6)   & 4.8 (0.5) & 58 (6)     & 21 (2) \\
ORIA-12\_02 & 05~35~12.3 & -06~37~30 & $1.8 (0.1) \times 10^{23}$  & 642 (64)   & 80 (8)   & 7.7 (0.8) & 253 (26)   & 53 (5) \\
ORIA-12\_01 & 05~35~23.9 & -06~58~30 & $1.9 (0.3) \times 10^{23}$  & 98 (10)    & 31 (3)   & 5.6 (0.6) & 25 (3)     & 11 (1)  \\
\hline
\noalign{\smallskip}
\end{tabular}
\tablefoot{A full version of this table is available at the CDS.
\tablefoottext{a}{The first number in the position name indicates the column density bin and the second one indicates the order in our sampling sequence.
Numbers in parentheses indicate 1$\sigma$ errors.}
        }      
\end{table*}

\clearpage

\section{Comparison with the Orion A data from \cite{yun21}}
\label{app_yun}

\begin{figure*}
        \centering
        \includegraphics[trim={5 290 5 60},clip,width=0.9\hsize]{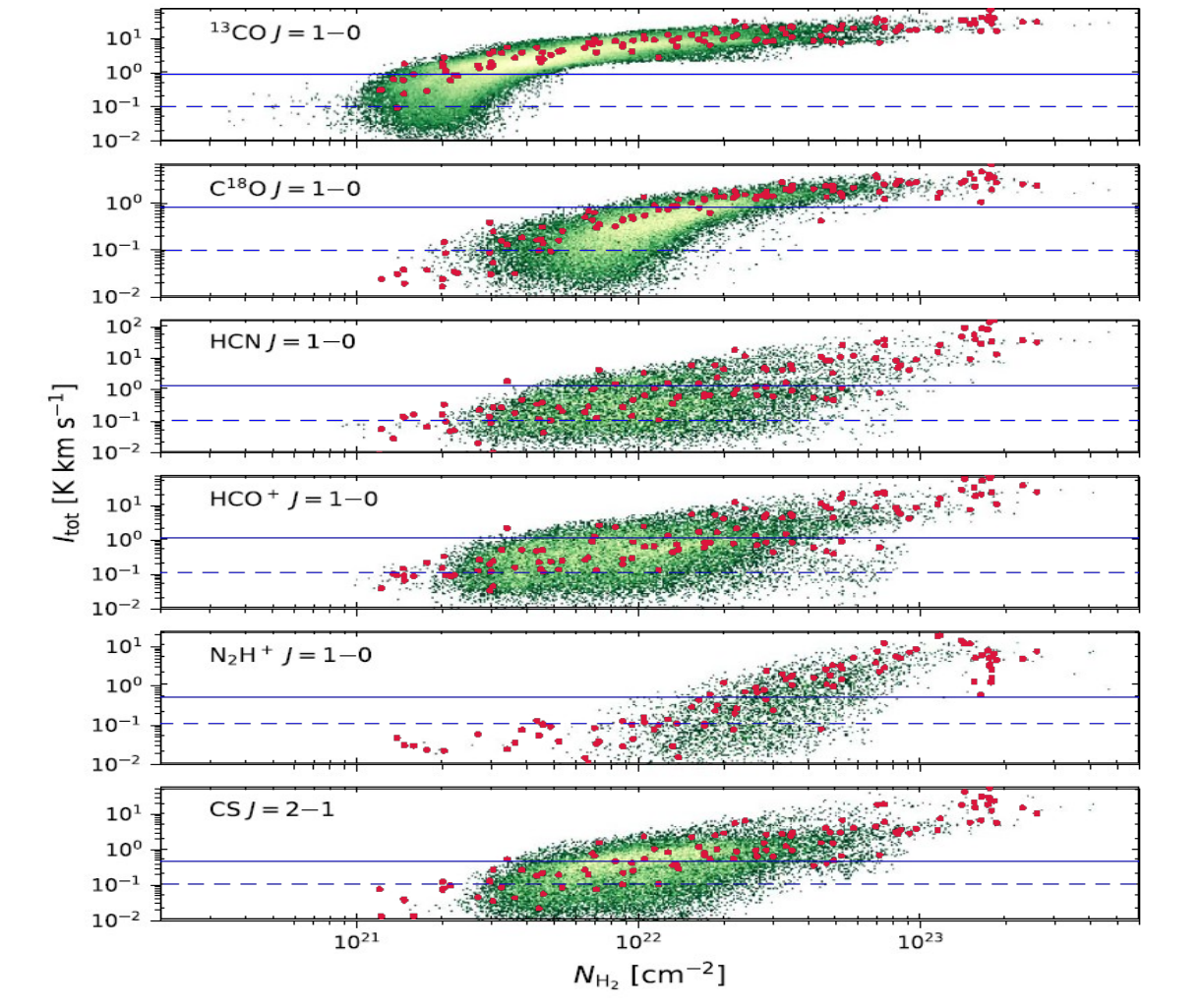}
        \caption{Comparison between the Orion A sampling observations and the
        mapping data of \cite{yun21}. The figure presents a scanned 
        copy of Fig.~15 from \cite{yun21}, with red circles representing the
        intensities derived from the sampling observations. The solid blue
        line represents the approximate $3 \sigma$ limit of the mapping data
        calculated using the velocity range
        of integration given by \cite{yun21} in their Table~2 and the rms
        noise in the spectra given in their Fig.~8. 
        The dashed line represents the equivalent limit for the sampling data.
        See the main text for further details.   
        }
        \label{fig_yun21}
\end{figure*}

Paper I presented a comparison between 
the Perseus $^{12}$CO(1--0) and $^{13}$CO(1--0) sampling data 
and the mapping observations of the same lines by
\cite{rid06} using the FCRAO 14 m telescope. This comparison
showed that sampling observations can reproduce at the 25\% level
both the mean intensity and the intensity
dispersion as a function of $N$(H$_2$) derived from the mapping 
observations (Paper I, Sect.~4.1).
For the Orion A cloud, \cite{yun21} have recently presented maps of several 
3 mm emission lines carried out with the 
twin TRAO
14 m telescope, and this
new data set provides an excellent opportunity to test again the ability of
the sampling method to reproduce the global emission of a molecular cloud.
The observations of \cite{yun21} consist of maps in the lines of $^{13}$CO(1--0), 
C$^{18}$O(1--0), HCN(1--0), CS(2--1), HCO$^+$(1--0), and N$_2$H$^+$(1--0) that
cover almost the full extent of the Orion A cloud emission (see the coverage in their 
Fig.~1). 
In contrast with the data from \cite{rid06}, the \cite{yun21} data are 
not publicly available, so our comparison with the Orion A data 
has been made by superposing our intensity distributions 
over those presented by \cite{yun21} in their Fig.~15.
Since \cite{yun21} presented their intensities in  
$T_\mathrm{A}^*$ units (Neal Evans, private communication), which 
correspond to
about half the intensities in units of $T_\mathrm{MB}$ (see the beam efficiencies
in their Table 1), to correctly 
simulate a line comparison in $T_\mathrm{MB}$ units
(as done with the data from \citealt{rid06}), we divided our
main-beam intensities by a factor of 2.
The results from the comparison are shown in
Fig.~\ref{fig_yun21}, which 
presents a superposition of our scaled-down intensities
(red circles) over the intensities presented by \cite{yun21} in their Fig.~15. 

As can be seen in Fig.~\ref{fig_yun21}, our sampling data match
well the  distribution of intensity as a function of 
$N$(H$_2$) derived from the mapping data for all observed lines. 
The best agreement is seen in the $^{13}$CO(1--0) data
since this line presents the highest signal to noise,
and its distribution can be reliably followed over 
two orders of magnitude in $N$(H$_2$).
The intensity of the other lines cannot be followed over such a 
large range of column densities, but in all cases, the sampling data 
overlap well with the distribution of mapping intensities 
and presents a similar level of dispersion over the
range of column densities for which the mapping data can be considered
complete (over the blue solid line).
Given the results in Fig.~\ref{fig_yun21}, we
conclude that the sampling observations provide a good
approximation to the distribution of intensities with $N$(H$_2$) 
found by mapping, and that they do not seem to miss any 
particular regime of intensities that is present in the maps.

A more quantitative comparison between our sampling data 
and the mapping observations of \cite{yun21} could be carried out using
line luminosities.
As discussed in our Sect.~\ref{sec_lum}, line luminosities can be easily
estimated from the sampling data by multiplying the observed line 
intensities by the surface area of the different column density bins,
and luminosity estimates for the main transitions observed in our three target
clouds are presented in Table.~\ref{tbl_luminosities}.
\cite{yun21} have also calculated luminosities for the lines observed in their
survey, but unfortunately the values given by these authors in their Sect.~5.4
suffer from several inaccuracies that make it necessary a revision 
(Neal Evans, private communication). A preliminary comparison with
the revised luminosities of $^{13}$CO(1--0) and C$^{18}$O(1--0)
suggests a level of agreement with our sampling results at
better than the 50\% level, which is 
similar to the agreement found in Sect.~\ref{sec_lum} when comparing our
luminosities with other estimates in the literature.
A definitive comparison  between our values
and those of \cite{yun21} awaits the final estimate of the
revised luminosities by these authors.

\clearpage
\clearpage

\section{Temperature effects}

\subsection{Gas temperature estimates}
\label{app_temp_est}

\begin{figure*}
        \centering
        \includegraphics[width=0.9\hsize]{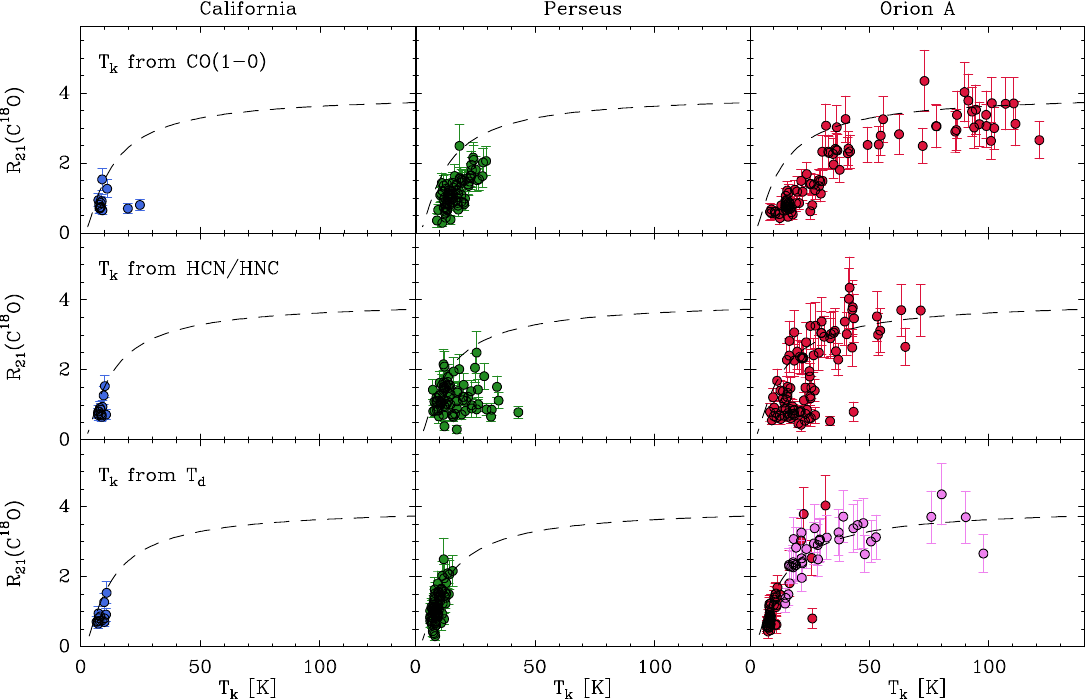}
        \caption{C$^{18}$O(2--1)/C$^{18}$O(1--0) intensity ratio as a function 
        of the gas temperature determined using three different methods:
        peak of the $^{12}$CO(1--0) line {\em (top),} HCN(1--0)/HNC(1--0) line
        ratio method of \cite{hac20} {\em (middle),} and the empirical
        fit based on the dust temperature described in the text {\em (bottom)}.
        The data are color-coded as in Fig.~\ref{fig_co1}, and the purple points in the 
        bottom-right panel indicate the Orion A positions where the NH$_3$-based temperature
        estimate of \cite{fri17} has been used.
        The dashed lines represent the expected line ratio under optically thin  LTE conditions,
        and the ability of the data to match this curve reflects the quality of the 
        temperature determination.
     }
        \label{fig_tk_3ways}
\end{figure*}

\begin{figure}
        \centering
        \includegraphics[width=0.9\hsize]{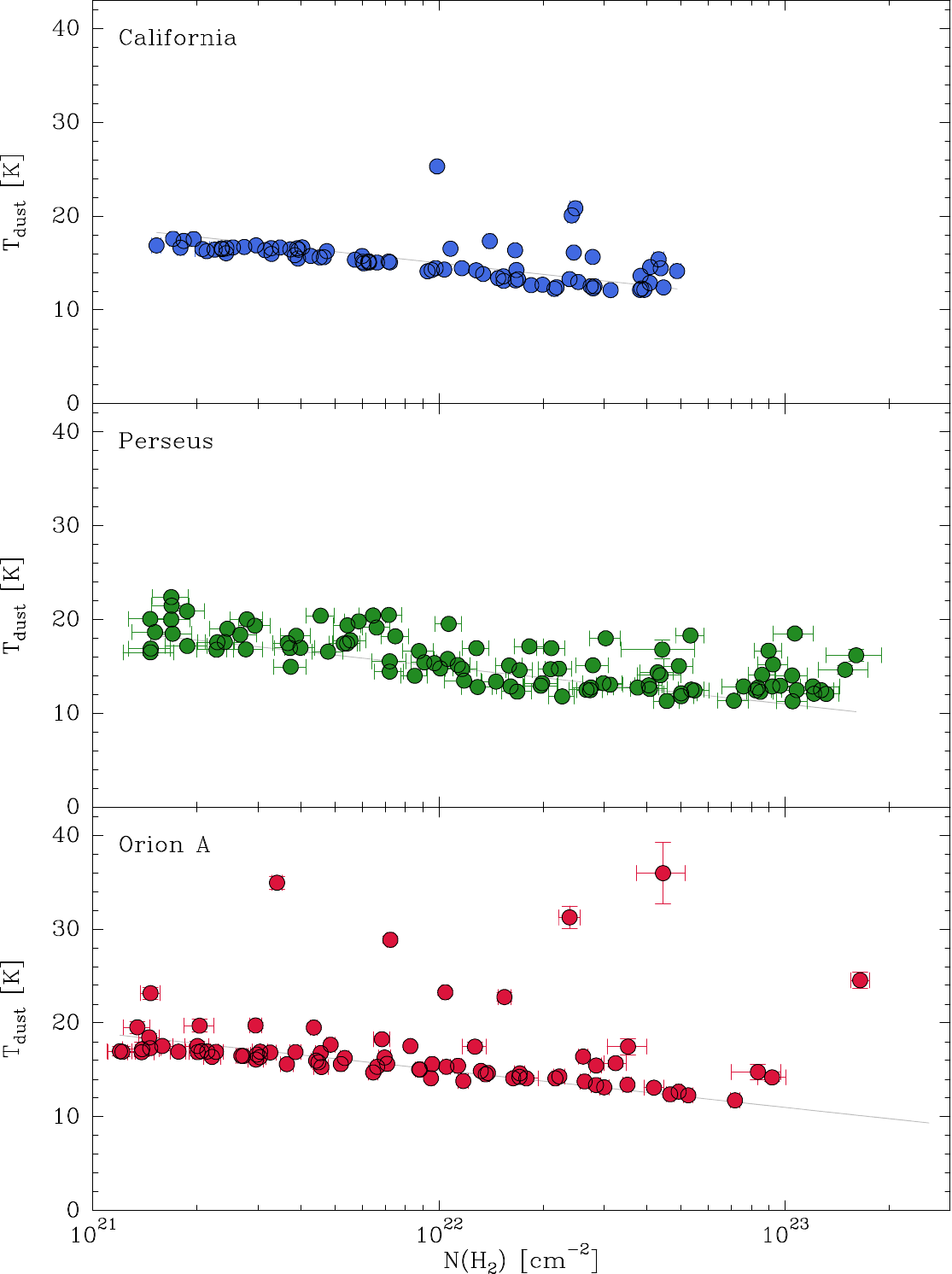}
        \caption{Distribution of dust temperature as a function of H$_2$ column
        density for the sampled positions of California, Perseus, and the regions of
        Orion A for which no NH$_3$-based gas temperature is available.
        The gray line represents
        the simple analytic fit to the equilibrium dust temperature described in the text.
        Dust temperature data are from \citet[California]{lad17}, \citet[Perseus]{zar16}, and \citet[Orion A]{lom14}.}
        \label{fig_tdust}
\end{figure}

To investigate the effect of temperature in the emission of the different lines, we
first needed to determine its mean value along each observed line of sight. 
The ideal tool for this is a tracer that is bright enough to
be detected at all column densities, has an excitation close to
the gas kinetic temperature, and is optically thin so it
samples the entire column density of gas at each position.
The tracer in our data set
that best approaches these characteristics is C$^{18}$O, and 
in Paper I we used the C$^{18}$O(2--1)/C$^{18}$O(1--0) ratio to determine 
that the temperature in the Perseus positions is approximately constant at about 11~K,
in agreement with previous ammonia results \citep{ros08}.
While a similar analysis could be done with the Orion A data, 
our C$^{18}$O(2--1) observations of California
were limited to 13 positions (out of 80) due to observing-time constraints.
As a result, it is not possible to derive a C$^{18}$O line ratio over most of this cloud,
and an alternative temperature determination is needed. In this appendix we 
explore three possible methods and judge their 
merit testing them against the C$^{18}$O line ratio,
which we consider the best available temperature indicator.

Our first (and simplest) temperature estimator is the peak intensity of
the $^{12}$CO(1--0) line. $^{12}$CO is easily
thermalized, and the $J$=1--0 line is optically thick.
Under these conditions, the gas kinetic temperature $T_\mathrm{k}$
can be assumed to equal the excitation temperature, which is given by 
\begin{equation}
T_\mathrm{ex} = \frac{h\nu/k}{\ln[1+\frac{h\nu/k}{T_\mathrm{B}-J_\nu(T_\mathrm{bg})}]},
\end{equation}
where $\nu$ is the line frequency, 
$J_\nu$ the radiation temperature, and 
$T_{\mathrm bg}$ the cosmic background temperature (=2.73~K).
To determine the $^{12}$CO(1--0) peak brightness temperature at each position, we fitted each spectrum using a Gaussian profile.
This type of profile does not always represent the best fit to the 
complex $^{12}$CO(1--0) spectrum, but a visual inspection of the fits shows that 
the method provides a good approximation to the line peak. Using this estimate and
Eq. D.1, we determined a $^{12}$CO(1--0)-based kinetic temperature 
at each cloud position in out sample. The top panel of
Fig.~\ref{fig_tk_3ways} presents the C$^{18}$O(2--1)/C$^{18}$O(1--0)
line ratio derived from our available data as a function of the gas temperature 
estimated using the  $^{12}$CO(1--0) peak temperature.
As can be seen, the line ratios for Perseus and Orion A
systematically lie to the right of
the dashed line, which corresponds to the prediction for the optically thin
local thermodynamic equilibrium (LTE) conditions expected to apply. This systematic shift 
indicates that the $^{12}$CO(1--0) peak method overestimates the value of the
C$^{18}$O-inferred
gas kinetic temperature by a significant margin.
A similar overestimate of the gas temperature by the $^{12}$CO data
was first noticed by  \cite{cas90} in their multiline CO study of 
Orion A, and recent discussions of this effect have been presented by \cite{hac20} 
and \cite{rou21}.
The most likely cause of this overestimate is the high optical depth of 
the $^{12}$CO(1--0) line, which
makes the emission originate from the cloud outer layers. These layers are 
exposed to the external UV radiation field and therefore warmer than the attenuated interior
(e.g., \citealt{cas90}).
A comparison of the CO-derived temperatures
in Orion A with the NH$_3$-based temperature estimates of \cite{fri17}
supports this interpretation, and suggests that the $^{12}$CO data
can overestimate the internal gas temperature by up to a factor of 2.

An alternative method for estimating the gas kinetic temperature is to use
the HCN(1--0)/HNC(1--0) line ratio, which behaves as a chemical thermometer due to the high 
sensitivity to temperature of the relative abundance of the two HCN isomers
\citep{hac20}.
Using NH$_3$ data from the Orion A cloud, \cite{hac20} have calibrated the relation 
between gas temperature and HCN/HNC ratio, and
proposed a simple prescription to derive the gas kinetic temperature (their Eqs. 3 and 4). 
We applied this prescription to our survey data, and determined a gas temperature
at each observed position. In the middle row of panels 
of Fig.~\ref{fig_tk_3ways} we present the C$^{18}$O(2--1)/C$^{18}$O(1--0) line ratio as
a function of the gas temperature estimated using the HCN/HNC line method.
As can be seen, the Orion A data follow more closely the 
optically thin LTE limit line than when using the $^{12}$CO(1--0) line method, although
this may simply reflect the fact that the HCN/HNC ratio was calibrated using NH$_3$ data from
this cloud. In contrast, the HCN/HNC ratio method does not align well the Perseus 
line ratios 
with the analytic curve, indicating that the method does not provide a reliable 
estimate of the gas kinetic temperature in this cloud. This should have been 
expected since the HCN/HNC ratio method  loses accuracy for temperatures
below 15~K \citep{hac20}.

As a third alternative to estimate gas temperatures, we explored the use of  
the dust temperature as a reference.
Dust temperatures are derived together with 
H$_2$ column densities in the analysis of the {\em Herschel} data 
\citep{lom14,zar16,lad17}, so they are available at each cloud position.
While the gas and dust components are only thermally coupled at 
densities in excess of $\approx 10^4$~cm$^{-3}$ \citep{gol01}, 
the dust temperature can still serve as an indicator of the presence (or absence) of 
energetic perturbations that may affect the gas temperature.
In our clouds, the largest temperature perturbation occurs in Orion A, where
the gas and dust in the vicinity of the ONC
are heated by the newly formed stars.
Since this temperature perturbation can be accurately characterized using the
NH$_3$-derived temperatures of \citep{fri17}, we used the NH$_3$ temperatures for 
the ONC vicinity, and used the dust temperature method only in the remaining parts of 
the clouds.
In these parts, the temperature excursions are likely very small, as 
suggested by the dust temperature profiles shown in Fig.~\ref{fig_tdust}.
These profiles indicate that the distribution of dust temperature in the three clouds has 
little dispersion, and that most values cluster along a well-defined band that runs 
almost horizontally. 
This band likely traces the equilibrium temperature of the
dust for unperturbed cloud conditions since
is very similar in the three clouds despite their significant
differences in star-formation activity.
It presents a slight decrease as $N$(H$_2$) increases that
likely results from the inward 
attenuation of the interstellar radiation field, which is
the main heating agent of the dust \citep{eva01,pla11}. 
Following work by \cite{hac17b} and \cite{hoc17},
we fitted the trend in the three clouds using the expression 
$T_\mathrm{dust} = 19.0 -4.0 \log_{10}(N(\mathrm{H}_2)/ 10^{21} \, \mathrm{cm}^{-2})$
(gray line in the figure).

To use the dust temperature as a proxy for the gas temperature, we followed 
a simple empirical approach. We first subtracted from
the dust temperature the observed gradient with $N$(H$_2$) since 
this gradient is caused by the attenuation of the interstellar radiation field,
and is not expected to affect the gas component because is 
heated by cosmic rays and not thermally 
coupled to the dust in the outer cloud \citep{gol01,gal02}.
To the result, we simply added a global offset of 8~K, which has been 
derived from an empirical match to the C$^{18}$O(2--1)/C$^{18}$O(1--0) line ratio.
As seen in the bottom row of panels in Fig.~\ref{fig_tk_3ways}, the 
empirical dust fit, when complemented with the NH$_3$-derived temperatures of \cite{fri17},
provides a better match to the C$^{18}$O(2--1)/C$^{18}$O(1--0) line ratio 
than the fits obtained using either 
the CO(1--0) peak intensity or the HCN/HNC line ratio.
For this reason, we used this fit as the temperature indicator in the
three clouds of our sample. Further work is needed to understand the correct
level of coupling between the gas and dust temperatures through the clouds.

\clearpage
\subsection{Temperature-correction factors}
\label{app_temp}

To determine (and possibly correct for) the effect of temperature variations in 
the emission profile of the different clouds, we needed to determine 
how the intensity of any line depends on the gas kinetic temperature. 
This determination
requires a certain degree of approximation because the intensity of a line
depends on a complex combination of the physical and chemical conditions
of the emitting cloud, which are in principle 
not well known. The observed systematic behavior 
of the line intensities, however,
suggests that the main characteristics of the emission
are dominated by relatively few cloud properties that could be approximated
using a simplified model. If this is the case, we could define a series of
correction factors that relate the intensity emitted by gas at an arbitrary 
temperature $T_{\mathrm k}$ with the emission that the same gas would emit
if it were a reference temperature of 10~K. These factors are defined as
\begin{equation}
f_{10\mathrm K} (T_{\mathrm k}) \equiv \frac{I\; (T_{\mathrm k})}
{I\; (10\mathrm{~K})},
\label{f_def}
\end{equation}
were $I(T_{\mathrm k})$ and $I(10~K)$ are the intensities at the two temperatures.
To first approximation, we could use these factors
to convert any observed line intensity into
the expected intensity that the same line of sight would emit if it were at 10~K.

As a first step to understand the behavior of the correction factors, 
we considered two simple limiting cases.
If the line is optically thick and thermalized ($T_{\mathrm ex} = T_{\mathrm k}$),
which may be the case for a low $J$ $^{12}$CO transition, the 
equation of radiative transfer states that the integrated
intensity will be equal to
$(J_\nu(T_{\mathrm k})-J_\nu(T_{\mathrm bg}))\; \Delta\mathrm{V}$, where
$J_\nu$ is the radiation temperature, 
$T_{\mathrm bg}$ is the cosmic background temperature, 
and $\Delta\mathrm{V}$ the linewidth in units of velocity.
Under these conditions, the correction factor defined as the
line intensity  at an arbitrary kinetic temperature divided by the intensity at 10~K 
will be
\begin{equation}
f^{\mathrm thick}_{10\mathrm K} (T_{\mathrm k}) = 
\frac{J_\nu(T_{\mathrm k})-J_\nu(T_{\mathrm bg})} {J_\nu(10~\mathrm{K})-J_\nu(T_{\mathrm bg})}.
\label{f_thick}
\end{equation}

\begin{figure*}
        \centering
        \includegraphics[width=\hsize]{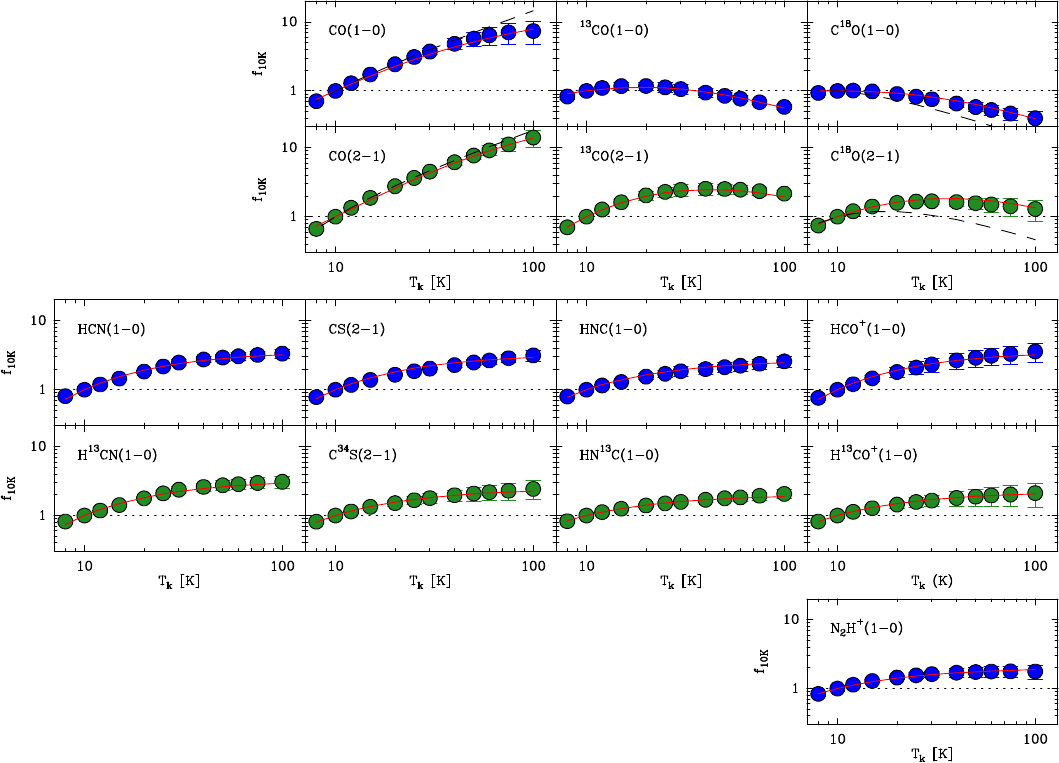}
        \caption{Intensity predictions of the observed lines
        as a function of gas temperature
        normalized to their value at 10~K. The intensities have been calculated
 by        running the model of the Perseus cloud presented in Paper I
        for 12 different gas temperatures.
        For each temperature, six different H$_2$ column densities 
        in the range $2\times 10^{21}$ to
        $2\times 10^{23}$~cm$^{-2}$ have been assumed, 
        and each circle represents the intensity average 
        with an error bar that indicates the 1$\sigma$ dispersion
        (and is often smaller than the marker size).
        The upper block of panels contains the CO isotopologs, the middle
        block contains the traditional
        dense-gas tracers and their isotopologs, and the bottom panel contains
        N$_2$H$^+$(1--0). The solid red lines represent the empirical
        fits described in the text and used in the analysis
        to correct for gas temperature variations. The dashed black lines in 
        the CO and C$^{18}$O panels represent the optically thick and thin 
        LTE solutions discussed in the text.    
        }
        \label{fig_tcorr}
\end{figure*}

Another simple illustrative case is the optically thin line of a species in LTE, in which the excitation temperature of all transitions
equals the kinetic temperature. For this case, which may be applicable to the
C$^{18}$O lines, the integrated
intensity will be $[J_\nu(T_{\mathrm k})-J_\nu(T_{\mathrm bg})]\; 
\tau_\nu\;\Delta\mathrm{V}$,
where the optical depth $\tau_\nu$ is given by
\begin{equation}
\tau_\nu = \frac{c^3}{{8 \pi \nu^3}}\; \frac{g_\mathrm{u} A_\mathrm{ul} N}{{\Delta V}}
\; \frac{(1-e^{{-h\nu}/kT_{\mathrm k}})\;e^{-E_{\mathrm{l}}/kT_\mathrm{k}}}{Q(T_\mathrm k)},
\end{equation}
and where $g_\mathrm{u}$ is the statistical weight of the upper level, 
$A_\mathrm{ul}$ is the Einstein coefficient of spontaneous emission, 
$N$ the molecular column density, and $Q$ the partition function
calculated as a sum over all relevant energy levels.
As in the thick case, we can derive a 
correction factor by dividing the expected line 
intensity at any given temperature
by the intensity at 10~K.

In real clouds, most molecular tracers are not thick and thermalized 
or thin and in LTE, 
so the conversion factors between temperatures will likely 
deviate from the previous two simple limits.
To calculate these factors, we therefore had to 
solve the equation of radiative transfer under physical and chemical
conditions that match those of real clouds.
In Paper I, we
presented a radiative transfer model of the Perseus cloud 
that was able to reproduce the 
intensity distribution of multiple lines (see 
Tables 3 and D.1 in Paper I ).
This model used a simple parameterization
of the gas physical and chemical conditions, and in particular, 
assumed a constant gas temperature of 
11~K as indicated by previous ammonia estimates \citep{ros08}.

To model how the intensity of the different lines is expected to vary with the
gas kinetic temperature, we
reran the Perseus model for 12 different temperatures that 
cover the range 8-100~K.
For each gas temperature, we estimated the emergent intensity of the different 
lines as a function of H$_2$ column density using
six logarithmically spaced H$_2$ column densities that cover the range
$2\times 10^{21}$-$2\times 10^{23}$~cm$^{-2}$. For each line,
we normalized the intensity at any given temperature
by the value at a reference temperature of 10~K
as described in Equ.~\ref{f_def}.

Figure~\ref{fig_tcorr} (top panels) 
shows the results from our radiative transfer model for the observed
transitions of the CO isotopologs. The circles represent the 
intensity ratio averaged over the range of column densities covered by each model,
and the error bars represent the corresponding dispersion. 
The red lines represent a set of analytic fits
described below, 
and the black dashed lines represent the two previously described limits: the 
optically thick thermalized limit for $^{12}$CO and the optically thin LTE
case for C$^{18}$O.
As can be seen, the model CO intensities approximately follow the 
optically thick prediction at low temperatures. 
As this parameter increases over about 50~K, however, the higher $J$
levels of CO become more populated and the optical depth of the lower
transitions starts to drop, decreasing the match with the thick approximation.
As also shown in the figure, the model C$^{18}$O intensities follow the optically 
thin LTE limit at low temperatures, but as the temperature increases, 
the LTE approximation breaks down and the intensities deviate from the 
dashed line.

Model results for the intensity ratio of the traditional dense-gas tracers and 
their isotopologs are
presented in the middle group of panels of Fig.~\ref{fig_tcorr}, and the 
special case of N$_2$H$^+$(1--0) is shown in the bottom panel.
As can be seen, the optically thick lines from the main isotopologs present a
higher sensitivity 
to the temperature increase, but the general behavior of both 
thick and thin lines is similar. It is characterized by 
an increase of the line ratio at low temperatures and 
a gradual flattening as the temperature exceeds 30-50~K.

We can summarize the results of our modeling by saying that the intensity of all
the observed lines is expected to vary smoothly and systematically with the gas 
kinetic temperature
in the range of interest (8-100~K).
Our model results provide a simple recipe to quantify and,
if necessary, correct for changes in the gas temperature across each cloud.
Since our model results have been calculated for a discrete number of
temperatures, it may be necessary to interpolate the corrections to the exact
temperature value of each position. This could be done by
interpolating the model results, but 
it seems more convenient to find
analytic expressions that fit the model results and 
can be used to quickly compute
correction factors for any gas temperature.

Inspired by the similar behavior of all the different dense-gas tracers, and by some of the factors that appear in the LTE equations, 
we explored the possibility of fitting the dense-gas tracer data 
using the simple expression 
\begin{equation}
 f_{10\mathrm K} (T_{\mathrm k}) = \frac{\exp (-T_0/T_{\mathrm{k}})}
 {\exp (-T_0/10\mathrm{~K})},
\end{equation}
where $T_0$ is a reference temperature that has been determined
by fitting the data in Fig.~\ref{fig_tcorr}. The fit results
are represented with red lines in the figure,
and the $T_0$ values for each dense gas transition are given in
Table~\ref{tbl_tcorr_dense}.
As can be seen, a simple exponential factor provides a reasonable fit to all 
dense-gas tracer lines 
and requires $T_0$ values in the narrow range 7-13~K.

Fitting the intensity ratios of the CO isotopologs requires more complex 
expressions because the model results present a diversity of behaviors
that range
from the quasi-linear increase of the thick main isotopologs to the systematic
decrease of the C$^{18}$O lines. Using again the LTE
formulas as an inspiration, and favoring the use of
similar factors for similar lines, we have found that it is possible to
fit the model results using an expression of the form
\begin{equation}
 f_{10\mathrm K} (T_{\mathrm k}) = \frac{g_1(T_\mathrm k)}{g_1(10\mathrm{~K})}
 \times \frac{g_2(T_\mathrm k)}{g_2(10\mathrm{~K})},
\end{equation}
where $g_1(T_\mathrm k)$ and $g_2(T_\mathrm k)$ are the empirical
functions given in Table~\ref{tbl_tcorr_co}.
As can be seen in Fig.~\ref{fig_tcorr}, this factorization provides a
reasonable fit to intensity correction factors of all
CO isotopologs in the 8-100~K range.
While convenient, these expressions only represent
empirical fits to the model results, and do not have any
underlying theoretical basis.

\begin{table}
\caption[]{Reference T$_0$ temperature for dense-gas tracers.
\label{tbl_tcorr_dense}}
\centering
\begin{tabular}{cc|cc}
\hline\hline
\noalign{\smallskip}
Line & $T_0$ [K] & Line & $T_0$ [K] \\
\noalign{\smallskip}
\hline
\noalign{\smallskip}
HCN(1--0) & 13 & H$^{13}$CN(1--0) & 12 \\
CS(2--1) & 12 & C$^{34}$S(2--1) & 9 \\
HNC(1--0) & 10 & HN$^{13}$C(1--0) & 7 \\
HCO$^+$(1--0) & 13 & H$^{13}$CO$^+$(1--0) & 8 \\
N$_2$H$^+$(1--0) & 7 & & \\
\noalign{\smallskip}
\hline
\noalign{\smallskip}
\end{tabular}
\end{table}

\begin{table*}
\centering
\caption[]{Temperature correction factors for CO isotopologs.
\label{tbl_tcorr_co}}
\begin{tabular}{ccc|ccc}
\hline\hline
\noalign{\smallskip}
Line & $g_1(T_\mathrm k)$ & $g_2(T_\mathrm k)$ & & $g_1(T_\mathrm k)$ & $g_2(T_\mathrm k)$ \\
\noalign{\smallskip}
\hline
\noalign{\smallskip}
CO(1-0) & $T_\mathrm{k}^{1.3}$ &  $(1-e^{-50/T_{\mathrm{k}}})$ &
CO(2-1) & $T_\mathrm{k}^{1.4}$ &  $(1-e^{-80/T_{\mathrm{k}}})$ \\
\noalign{\smallskip}
$^{13}$CO(1-0) & $e^{-4/T_{\mathrm{k}}}$ &  $(1-e^{-50/T_{\mathrm{k}}})$ &
$^{13}$CO(2-1) & $e^{-14/T_{\mathrm{k}}}$ &  $(1-e^{-80/T_{\mathrm{k}}})$ \\
\noalign{\smallskip}
C$^{18}$O(1-0) & 1 & $(1-e^{-50/T_{\mathrm{k}}})$ &
C$^{18}$O(2-1) & $e^{-10/T_{\mathrm{k}}}$ &  $(1-e^{-80/T_{\mathrm{k}}})$ \\
\noalign{\smallskip}
\hline
\noalign{\smallskip}
\end{tabular}
\end{table*} 
 
\clearpage

\section{Plots of isotopic ratios}
\label{app_isotop}

\begin{figure}[h]
        \centering
        \includegraphics[width=\hsize]{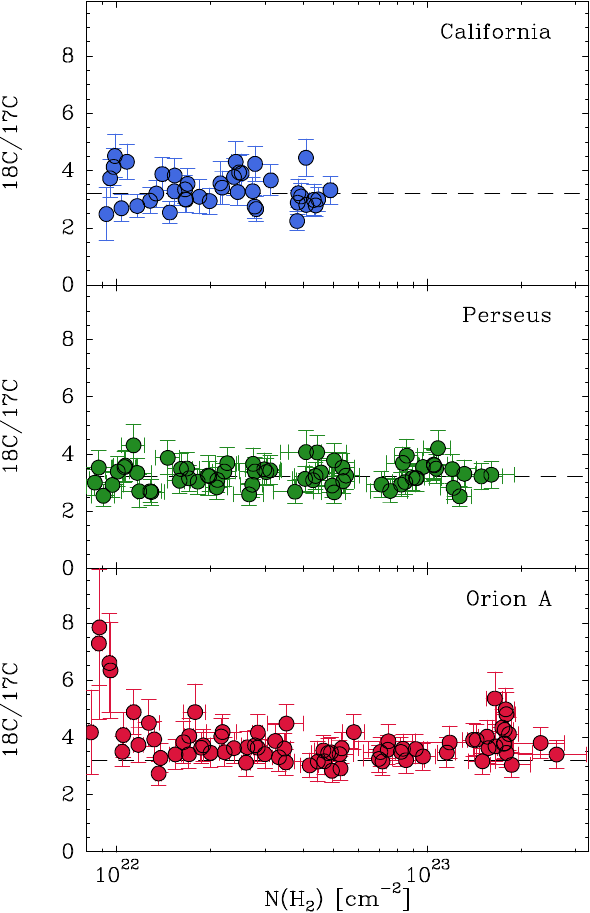}
        \caption{C$^{18}$O(1--0)/C$^{17}$O(1--0) intensity ratio as a function of 
        the H$_2$ column density used to determine the optical depth of the rare 
        isotopologs. The dashed horizontal line represents the standard isotopic 
        ratio of 3.2 \citep{wil94}, and the good match with the observations
        indicates that both transitions are optically thin in all three clouds. 
        Note the restricted
        range of the x-axis due to the low S/N of the line ratio for very low
        values of $N$(H$_2$).}
        \label{fig_c18o_c17o}
\end{figure}

\begin{figure}
        \includegraphics[width=\hsize]{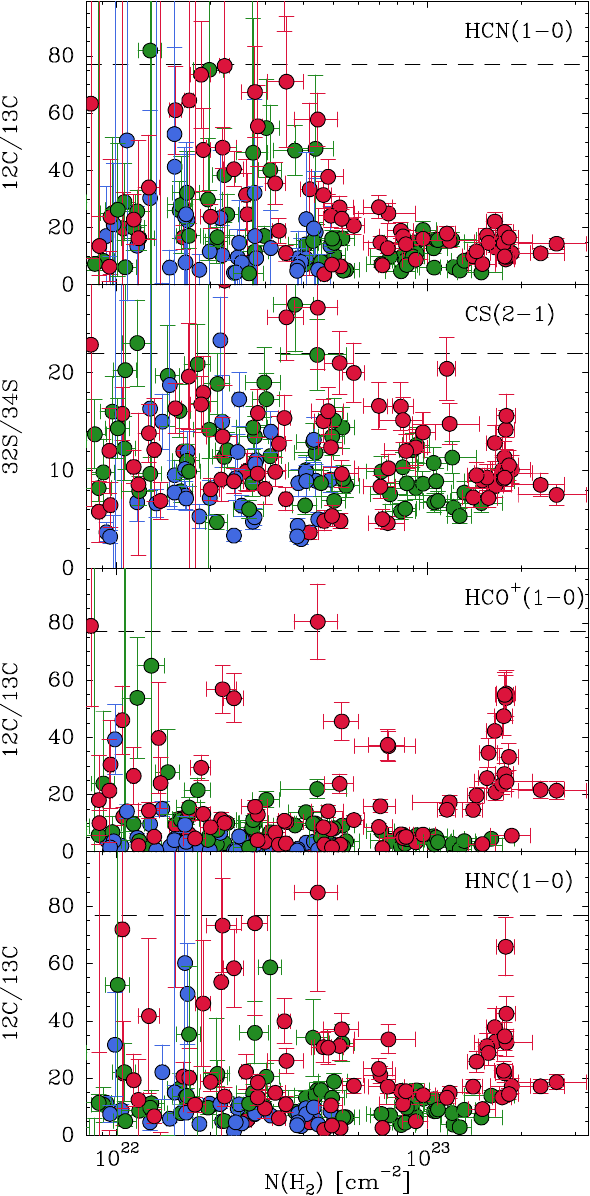}
\caption{Isotopic intensity ratios for the traditional dense-gas tracers
as a function of H$_2$ column density for California, Perseus, and Orion A.
The $N$(H$_2$) range has been reduced compared to other plots due to the
low signal-to-noise ratio of the data with low $N$(H$_2$). The dashed horizontal 
lines represent the standard interstellar medium isotopic ratios
(77 for $^{12}$C/$^{13}$C 
and 22 for $^{32}$S/$^{34}$S), and the low value of the observed ratios
indicates that the emission of the main isotopolog is optically thick. 
Note how the outflow points from Orion A (near $2\times 10^{23}$~cm$^{-2}$)
have lower optical depths in some tracers but are still optically thick.
No temperature correction has been applied to the data.}
        \label{fig_dense_ratios}
\end{figure}

Figs.~\ref{fig_c18o_c17o} and \ref{fig_dense_ratios} present intensity ratios
as a function of H$_2$ column density for the rare CO isotopologs and the
traditional dense-gas tracers. A comparison with the standard
isotopic value (represented with a dashed line) 
indicates that the emission from the rare 
CO isotopologs is optically thin, while the emission from  
the dense-gas tracers is optically thick.

\clearpage

\section{Luminosities from mapping and sampling methods}

Table \ref{tbl_luminosities_comparison} presents the set of 
luminosity values used in the 
comparison between sampling and mapping luminosities shown in 
Fig.~\ref{fig_luminosity_test}. The luminosities from mapping observations
were chosen from the literature requiring that they were well documented
and reliable calibrated, or that the mapping data were publicly available so that
the luminosities could be independently estimated. If necessary, luminosities have
been re-scaled to the \textit{Gaia}-based distances of \cite{zuc19} given in
Sect.~\ref{sec_intro}. To compare with the Perseus estimates from \cite{rid06}
and the Orion A estimates from \cite{nis15}, we restricted our sampling
estimates to the same cloud area not masked by these authors, for which 
we used their
publicly available data. A full 
discussion of the 
luminosity comparison is given in Sect.~\ref{sec_lum}.

\begin{table}[h]
\caption[]{Line luminosities from mapping and sampling measurements.
\label{tbl_luminosities_comparison}}
\centering
\begin{tabular}{lccc}
\hline\hline
\noalign{\smallskip}
Line & $L_\mathrm{sampling}$ & $L_\mathrm{mapping}$ & 
Map ref.\tablefootmark{a} \\
 & (K km s$^{-1}$ pc$^2$) & (K km s$^{-1}$ pc$^2$) \\
\noalign{\smallskip}
\hline
\noalign{\smallskip}
\multicolumn{4}{c}{Orion A} \\
\noalign{\smallskip}
\hline
\noalign{\smallskip}
$^{12}$CO(1--0) & 28,200 & 21,300 & Lew22 \\
$^{12}$CO(2--1) & 19,100 & 16,000 & Nis15 \\
$^{13}$CO(2--1) & 3,350 & 2,030 & Nis15 \\
C$^{18}$O(2--1) & 91 & 53 & Nis15 \\
\noalign{\smallskip}
\hline
\noalign{\smallskip}
\multicolumn{4}{c}{Perseus} \\
\noalign{\smallskip}
\hline
\noalign{\smallskip}
$^{12}$CO(1--0) & 3,890 & 3,920 & Rid06 \\
$^{12}$CO(1--0) & 6,900 & 8,530 & Lew22 \\
$^{13}$CO(1--0) & 700 & 810 & Rid06 \\
HCN(1-0) & 77 & 67 & Dam23 \\
\noalign{\smallskip}
\hline
\noalign{\smallskip}
\multicolumn{4}{c}{California} \\
\noalign{\smallskip}
\hline
\noalign{\smallskip}
$^{12}$CO(1--0) & 35,200 & 30,500 & Lew22 \\
\noalign{\smallskip}
\hline
\end{tabular}
\tablefoot{
\tablefoottext{a}{Reference code for mapping luminosities: 
Dam23 = \cite{dam23}, Lew22 = \cite{lew22}, 
Nis15 = \cite{nis15}, Rid06 = \cite{rid06}}
}
\end{table}

\end{document}